\documentclass[pdflatex, twocolumn, sn-nature]{sn-jnl}% Math and Physical Sciences Numbered Reference Style
\usepackage{graphicx}%
\usepackage{multirow}%
\usepackage{amsmath,amssymb,amsfonts}%
\usepackage{amsthm}%
\usepackage{mathrsfs}%
\usepackage[title]{appendix}%
\usepackage{xcolor}%
\usepackage{textcomp}%
\usepackage{manyfoot}%
\usepackage{booktabs}%
\usepackage{algorithm}%
\usepackage{algorithmicx}%
\usepackage{algpseudocode}%
\usepackage{listings}%

\providecommand\aj{Astron.\ J.} % Astronomical Journal
\providecommand\araa{Annu. Rev. Astron. Astrophys.} % Annual Review of Astron and Astrophys
\providecommand\aap{Astron.\ Astrophys.} % Astron and Astrophys
\providecommand\aaps{Astron.\ Astrophys.\ Suppl.} % Astronomy and Astrophysics Supplement
\providecommand\apj{Astrophys.\ J.} % Astrophysical Journal
\providecommand\apjl{Astrophys.\ J.} % Astrophysical Journal, Letters
\providecommand\apjs{Astrophys.\ J.\ Suppl.} % Astrophysical Journal, Supplement
\providecommand\icarus{Icarus} % Icarus
\providecommand\pasp{Publ. Astron. Soc. Pac.} % Publications of the ASP
\providecommand\nat{Nature} % Nature
\providecommand\jqsrt{J.\ Quant.\ Spectrosc.\ Radiat.\ Transfer} % J. Quant. Spectrosc. Radiat. Transfer
\providecommand\mnras{Mon.\ Not.\ R.\ Astron.\ Soc.} % Monthly Notices of the RAS
\providecommand\planss{Planet.~Space~Sci.} % Planetary Space Science
\providecommand\psj{Planet.\ Sci.\ J.} % The Planetary Science Journal
\providecommand\ssr{Space~Sci.~Rev.} % Space Science Reviews
\providecommand\nar{New Astronomy Review}
\providecommand\grl{Geophysics Research Letters}
\theoremstyle{thmstyleone}%
\theoremstyle{thmstyletwo}%

\theoremstyle{thmstylethree}%

\definecolor{mygreen}{RGB}{0,128,0}

\begin{document}

\title[Water D/H in 3I/ATLAS as a Probe of Formation Conditions in Another Planetary System]{Water D/H in 3I/ATLAS as a Probe of Formation Conditions in Another Planetary System}

\author*[1]{\fnm{Luis E.} \sur{Salazar Manzano}}\email{lesamz@umich.edu}\equalcont{These authors contributed equally to this work.}

\author[1]{\fnm{Teresa} \sur{Paneque-Carre\~no}}\equalcont{These authors contributed equally to this work.}

\author[2,3]{\fnm{Martin A.} \sur{Cordiner}}

\author[1]{\fnm{Edwin A.} \sur{Bergin}}

\author[4,5]{\fnm{Hsing Wen} \sur{Lin}}

\author[6]{\fnm{Dariusz C.} \sur{Lis}}

\author[7,4,1]{\fnm{David W.} \sur{Gerdes}}

% After Gerdes (alphabetical by surname)
\author[8]{\fnm{Jennifer B.} \sur{Bergner}}

\author[9]{\fnm{Nicolas} \sur{Biver}}

\author[9]{\fnm{Dominique} \sur{Bockel\'ee-Morvan}}

\author[10]{\fnm{Dennis} \sur{Bodewits}}

\author[2]{\fnm{Steven B.} \sur{Charnley}}

\author[9]{\fnm{Jacques} \sur{Crovisier}}

\author[6]{\fnm{Davide} \sur{Farnocchia}}

\author[11,12]{\fnm{Viviana V.} \sur{Guzm\'an}}

\author[2]{\fnm{Stefanie N.} \sur{Milam}}

\author[10]{\fnm{John W.} \sur{Noonan}}

\author[13]{\fnm{Anthony J.} \sur{Remijan}}

\author[2,14]{\fnm{Nathan X.} \sur{Roth}}

\author[13]{\fnm{John J.} \sur{Tobin}}

% --- Affiliations ---

% 1. Michigan Astro
\affil*[1]{\orgdiv{Department of Astronomy}, \orgname{University of Michigan},
  \orgaddress{\city{Ann Arbor}, \state{MI}, \postcode{48109}, \country{USA}}}

% 2. NASA Goddard
\affil[2]{\orgdiv{Astrochemistry Laboratory}, \orgname{NASA Goddard Space Flight Center},
  \orgaddress{\street{8800 Greenbelt Road}, \city{Greenbelt}, \state{MD}, \postcode{20771}, \country{USA}}}

% 3. Catholic Univ
\affil[3]{\orgdiv{Department of Physics}, \orgname{The Catholic University of America},
  \orgaddress{\city{Washington}, \state{DC}, \postcode{20064}, \country{USA}}}

% 4. Michigan Physics
\affil[4]{\orgdiv{Department of Physics}, \orgname{University of Michigan},
  \orgaddress{\city{Ann Arbor}, \state{MI}, \postcode{48109}, \country{USA}}}

% 5. MIDAS
\affil[5]{\orgdiv{Michigan Institute for Data and AI in Society}, \orgname{University of Michigan},
  \orgaddress{\city{Ann Arbor}, \state{MI}, \postcode{48109}, \country{USA}}}

% 6. JPL
\affil[6]{\orgdiv{Jet Propulsion Laboratory}, \orgname{California Institute of Technology},
  \orgaddress{\street{4800 Oak Grove Drive}, \city{Pasadena}, \state{CA}, \postcode{91109}, \country{USA}}}

% 7. Case Western
\affil[7]{\orgdiv{Department of Physics}, \orgname{Case Western Reserve University},
  \orgaddress{\city{Cleveland}, \state{OH}, \postcode{44106}, \country{USA}}}

% 8. Berkeley
\affil[8]{\orgdiv{Department of Chemistry}, \orgname{University of California, Berkeley},
  \orgaddress{\city{Berkeley}, \state{CA}, \country{USA}}}

% 9. Paris
\affil[9]{\orgname{LIRA, Observatoire de Paris, Universit\'e PSL, CNRS, Sorbonne Universit\'e, Universit\'e Paris Cit\'e},
  \orgaddress{\street{5 place Jules Janssen}, \city{Meudon}, \postcode{92195}, \country{France}}}

% 10. Auburn
\affil[10]{\orgdiv{Physics Department}, \orgname{Edmund C. Leach Science Center, Auburn University},
  \orgaddress{\city{Auburn}, \state{AL}, \postcode{36849}, \country{USA}}}

% 11. Chile (PUC)
\affil[11]{\orgdiv{Instituto de Astrof\'isica}, \orgname{Pontificia Universidad Cat\'olica de Chile},
  \orgaddress{\street{Av. Vicu\~na Mackenna 4860, Macul}, \city{Santiago}, \postcode{7820436}, \country{Chile}}}

% 12. YEMS (New affiliation for Guzman, replacing the slot left by Edinburgh)
\affil[12]{\orgname{Millennium Nucleus on Young Exoplanets and their Moons (YEMS)},
  \orgaddress{\city{Santiago}, \country{Chile}}}

% 13. NRAO
\affil[13]{\orgname{National Radio Astronomy Observatory},
  \orgaddress{\street{520 Edgemont Rd}, \city{Charlottesville}, \state{VA}, \postcode{22903}, \country{USA}}}

% 14. American Univ
\affil[14]{\orgdiv{Department of Physics}, \orgname{American University},
  \orgaddress{\street{4400 Massachusetts Avenue, NW}, \city{Washington}, \state{DC}, \postcode{20016}, \country{USA}}}

% \cite{1981A&A....93..189G,1989ApJ...340..906M,Alexander:2018SSRv}
% \cite{Ceccarelli:2014prpl.conf,Cleeves:2014Sci, Nomura:2023ASPC, Ciesla:2026AJ}
% \cite{Bergin:2024come.book....3B}
\abstract{
Water reservoirs in the Solar System exhibit a deuterium enrichment that links back to the physical environment at the time of stellar birth. Gas-phase and ice-grain deuterium enrichments occur through chemical processes that operate at low temperatures ($<$~30~K) pointing towards an origin in the prestellar molecular cloud or in the outer parts of the protoplanetary disk. However, not all stars are born in environments similar to our Sun, nor do their subsequent evolutionary histories follow the same path. These environmental differences can be traced by the water deuterium-to-hydrogen (D/H) ratio. Here we use ALMA observations of the interstellar comet 3I/ATLAS to constrain the water D/H ratio in extrasolar cometary material. With a water D/H value of [D/H]$_{\mathrm{H_2O}} > 6.6\times10^{-3}$, 3I/ATLAS shows a deuterium enrichment exceeding Earth's ocean value by more than a factor of $\gtrsim40$ and typical Solar System cometary values by more than a factor of $\gtrsim30$. The elevated deuterium enrichment points to water that formed under colder, less irradiated conditions and from less thermally processed material, consistent with an origin in a planetary system that formed under different physical and chemical conditions than our own.
    }

\keywords{Interstellar Comets, Comets, Astrochemistry, Comet Volatiles}



\maketitle

% \section{Main Text}\label{sec1}

Among the vast diversity of chemical species, water stands out as an essential molecule for life and astrophysical processes \cite{vanDishoeck:2013ChRv, Westall:2018SSRv}. From an astrobiology perspective, water is a key solvent for the emergence of life on Earth and is traced throughout the Universe as a potential signpost of habitable environments beyond our planet \cite{Chyba:2005ARA&A}. In the context of star and planet formation, water in the gas phase acts as an efficient coolant, allowing molecular clouds to collapse into forming stars. In frozen form, water coats dust grains, allowing them to stick together more efficiently and enabling the rapid growth of planetary cores \cite{vanDishoeck2014prpl.conf}. Water has been detected in both the gas and ice phases throughout our Galaxy and in high-redshift galaxies. These detections span molecular clouds, protostar systems, prestellar cores, protoplanetary disks, and Solar System bodies, including comets, meteorites, active asteroids, planets, and satellites \cite{Encrenaz:2008ARA&A, Bergin:2012RSPTA, Kelley:2023Natur, Brown:2025PSJ}. Current studies aim to connect the path of water across these diverse environments to understand its  origin and evolution in forming planetary systems \cite{Ceccarelli:2014prpl.conf}. 

The deuterium-to-hydrogen (D/H) ratio in water provides a powerful chemical tracer of where water formed, the physical conditions under which it originated, and how it was subsequently processed. Deuterium fractionation,  the process that enriches the abundance of D relative to H in molecules \cite{Ceccarelli:2014prpl.conf}, is highly sensitive to local conditions including temperature, density, and ionization rate. In cold environments such as prestellar molecular cores, fractionation is dominated by the gas phase reaction $\mathrm{H_3^+ + HD \rightleftharpoons H_2D^+ + H_2 + \Delta E}$, where $\Delta E$ denotes the energy barrier for the reverse reaction. As a result, the forward reaction is favored at low temperatures, enhancing the production of $\mathrm{H_2D^+}$ \cite{Roberts:2003ApJ}. Subsequent dissociative recombination enhances the abundance of atomic D, leading to elevated gas-phase atomic D/H ratios that become incorporated into water ice \cite{Tielens:1983A&A}. This chemical signature can survive through later evolutionary stages of the material \cite{Ceccarelli:2014prpl.conf, Leemker:2025NatAs}, but it may be partially erased during warmer protostellar and protoplanetary phases through molecular dissociation or backward isotope exchange reactions. Observationally, constraining the water D/H ratio is difficult because the low-energy H$_2$O bands suffer from strong atmospheric absorption, and because HDO is intrinsically much less abundant than H$_2$O. Extending these constraints to physically and chemically distinct regions of our Galaxy is even more challenging because of their large distances.

Comets are icy planetesimals, preserved relics of the planet formation process \cite{AHearn:2011ARA&A}. Their ice composition tends to be dominated by water \cite{Bockelee-Morvan:2017RSPTA}, making them powerful tracers of the chemical composition and physical conditions under which water formed \cite{Mumma:2011ARA&A, Dones:2015SSRv, Guilbert-Lepoutre:2015SSRv, BockeleeMorvan:2015SSRv}. Comets in other planetary systems in the Milky Way have long been expected to exist, and observations of exocomets (through dust or gas signatures, \cite{Ferlet:1987A&A}) together with white dwarf atmospheric pollution \cite{Farihi:2016NewAR} provide indirect evidence for such reservoirs \cite{Strom:2020PASP, Iglesias:2025SSRv}. Dynamical processes that operate in planetary systems, including scattering by giant planets, can eject planetesimals onto unbound trajectories, producing a population of interstellar objects \cite{Forbes:2025ApJ}. When an interstellar comet passes close to the Sun on a hyperbolic trajectory, sublimation of its ices \cite{Bodewits:2024come.book, Biver:2024come.book} provides direct access to material originating around other stars \cite{Jewitt:2023ARA&A, Fitzsimmons:2024come.book}.  The first interstellar object with confirmed outgassing, 2I/Borisov, was detected in 2019 \cite{Borisov:2019CBET}. In July 2025, 3I/ATLAS \cite{Denneau:2025} became the second interstellar visitor with a clearly resolved gaseous coma \cite{Jewitt:2025ATel, Alarcon:2025ATel, Seligman:2025ApJ}.

Since its discovery, 3I/ATLAS has shown distinctive properties that rapidly separated it not only from known Solar System comets but also from the previously known interstellar objects. Its estimated kinematic age of 3-11 Gyr (refs. \cite{Taylor:2025ApJ, Hopkins:2025ApJ}) suggests that it may be the oldest interstellar object discovered so far, probably formed in the early Galaxy. James Webb Space Telescope (JWST) and Spectro-Photometer for the History of the Universe, Epoch of Reionization and Ices Explorer (SPHEREx) near-infrared observations showed that the comet's abundance of carbon dioxide (CO$_2$) with respect to H$_2$O pre-perihelion was enhanced compared with Solar System comets at similar heliocentric distances \cite{Cordiner:2025ApJ, Lisse:2025RNAAS}. Atacama Large Millimeter/submillimeter Array (ALMA) submillimeter observations indicated a pronounced enrichment of methanol (CH$_3$OH) compared with HCN (ref. \cite{Roth:2026ApJ}). Optical spectroscopy obtained during its inbound leg revealed signatures of carbon-chain depletion \cite{SalazarManzano:2025ApJ, Schleicher:2025ATel, Hutsemekers:2026A&A} and atypical ratios of nickel relative to iron \cite{Rahatgaonkar:2025ApJ}, both of which appeared to evolve towards more typical values after perihelion \cite{Hutsemekers:2026A&A, Jehin:2025ATel}. As an object with a formation and evolutionary history that is probably very different from the Solar System comets, measurements of deuterated water in 3I/ATLAS offer a rare opportunity to probe D fractionation and initial prestellar conditions from a broader Galactic perspective. 

\begin{figure*}
    \centering
    \includegraphics[width=1\linewidth]{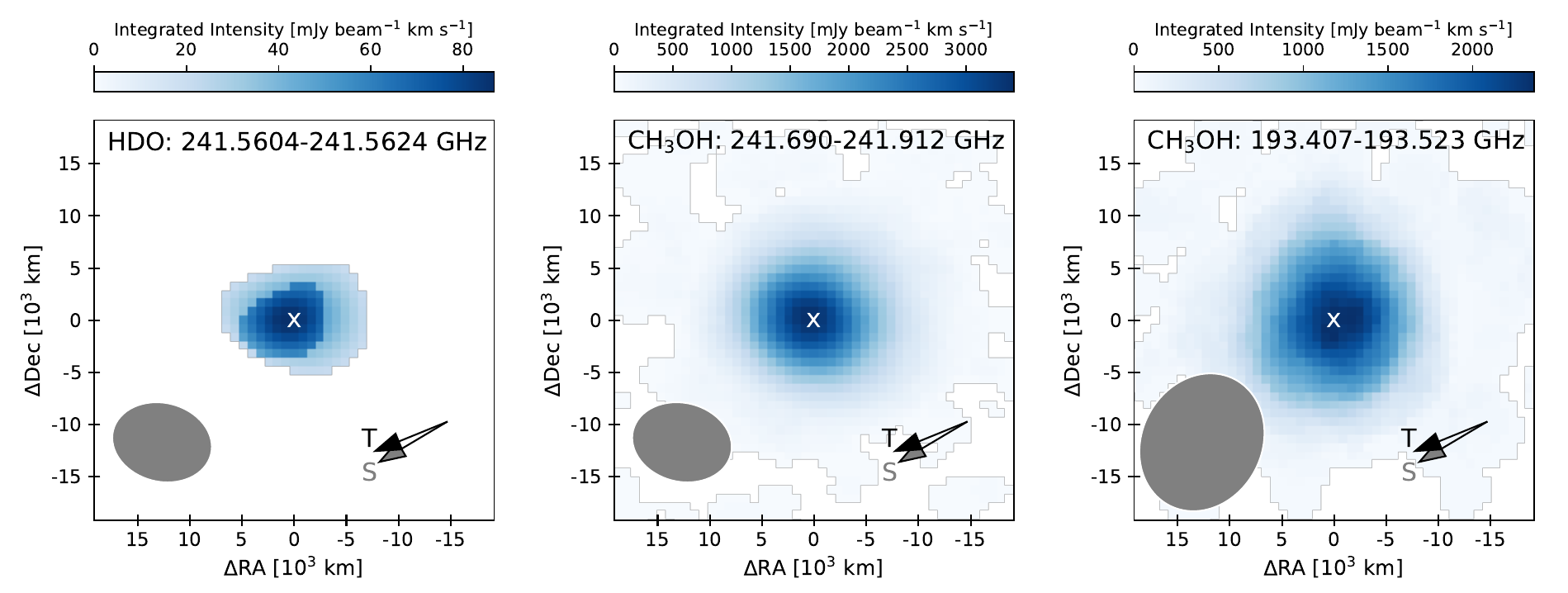}
    \caption{\textbf{Integrated intensity (moment 0) maps of the detected species}. From left to right: HDO, CH$_3$OH in Band 6 and CH$_3$OH in Band 5. For each image, the ellipse in the bottom left shows the size of the ALMA beam and in the bottom right, the direction of the Sun (S) and comet trail (T) are signaled with the corresponding arrows. The white cross marks the location of the comet nucleus. Dec, declination; RA, right ascension.}
    \label{fig:moment0_maps}
\end{figure*}

In this work, we constrain the water D/H in 3I/ATLAS from ALMA observations. The data were obtained through the ALMA Director Discretionary Time (DDT) program 2025.A.00002.T (principal investigator T.P.-C.; Extended Data Table~1). We targeted ALMA Band 5 and Band 6 spectral windows centered on H$_2$O ($J_{K_{a},K_{c}}$ = 3$_{1,3}$ -- 2$_{2,0}$ at $\nu$ = 183.310\,GHz) and HDO ($J_{K_{a},K_{c}}$ = 2$_{1,1}$ -- 2$_{1,2}$ at $\nu$ = 241.561\,GHz), respectively, together with multiple CH$_3$OH lines in Band 5 ($J_K=4_K-3_K$, 193 GHz) and Band 6 ($J_K=5_K-4_K$, 242 GHz). All observations were executed with the Atacama Compact Array (ACA), which combines a compact 7-m interferometric array with 12-m total-power antennas to improve sensitivity to emission on large angular scales.

Quasi-simultaneous observations of both ALMA Bands were executed on 4 November 2025. On that date, 3I/ATLAS was at a heliocentric distance, $r_H$, of 1.37 au and a geocentric distance, $\Delta$, of 2.24 au. This corresponds to 6 days after perihelion, which occurred on 29 October 2025 at $r_H$ = 1.36 au. The data were calibrated with the standard ALMA pipeline, continuum subtracted, and imaged with natural weighting to optimize sensitivity. The resulting root-mean-square (RMS) noise level in each 244.141\,kHz channel in the Band 6 CH$_3$OH and HDO dataset is 21~mJy/beam. For Band 5, the line centered spectral windows have a channel spacing of 122.070\,kHz and an RMS noise level of 500~mJy/beam for H$_2$O and 55~mJy/beam for CH$_3$OH. Further details of the set of observations can be found in Methods. HDO and multiple CH$_3$OH lines are detected, whereas H$_2$O remains below the detection threshold. Figure~1 shows the integrated emission maps of each detected molecule, calculated from emission above 3 RMS within a frequency range shown at the top of each panel.

To determine the physical and chemical properties of the 3I/ATLAS coma, we modeled the spectra extracted at the nucleus position, as described in Methods. Our modeling used the non-local thermodynamic equilibrium (non-LTE) radiative transfer code SUBLIME \cite{Cordiner:2022ApJ, Cordiner:2023ApJ} for each of our molecular tracers. The code computes the collisionally and radiatively excited molecular-level populations in a time-dependent, expanding cometary atmosphere and predicts the corresponding spectral line profiles. For this work, we used the one-dimensional (1D) implementation of SUBLIME, which assumes an isothermal and spherically symmetric outflow. As the modeling includes a Doppler offset $\Delta v$, the inferred kinematics represent an average towards and away from the observer's line of sight. However, given the relatively small phase angle at the time of observations ($\sim16^\circ$), the kinematics can be approximated as an average over the sunward and anti-sunward directions. Previous work has demonstrated that, for ACA ALMA observations of 3I/ATLAS, 1D SUBLIME modeling produces production rates consistent with three-dimensional (3D) SUBLIME models \cite{Roth:2026ApJ}, largely due to the optically thin nature of the observed emission.

\begin{figure*}
    \centering
    \includegraphics[width=1\linewidth]{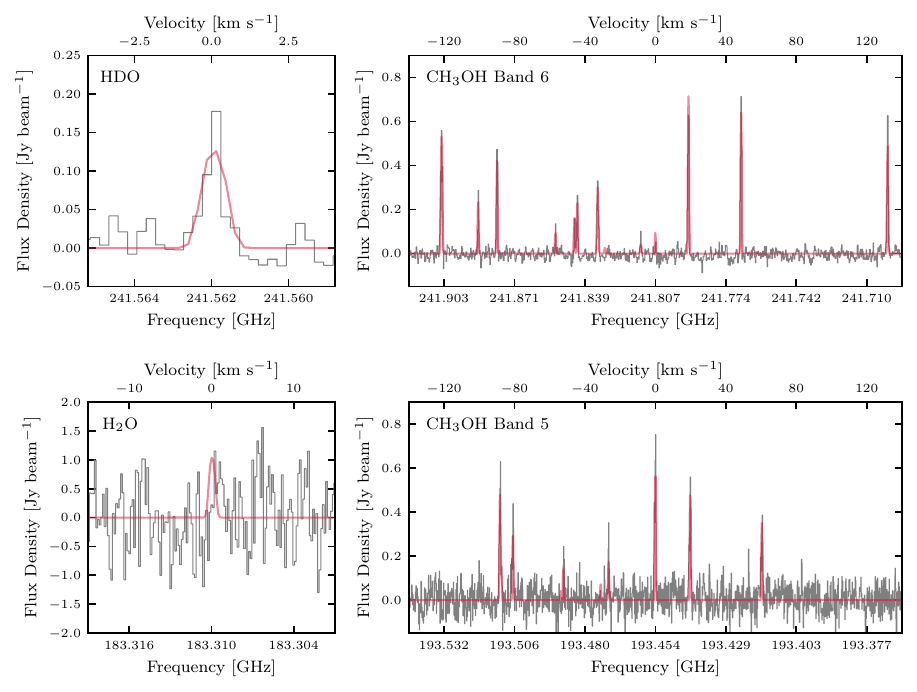}
    \caption{\textbf{3I/ATLAS ALMA spectra and best-fit models for HDO, H$_2$O, and CH$_3$OH in Bands 5 and Band 6}. The 3I/ATLAS spectra (gray) were extracted at the nucleus position, and the HDO, H$_2$O, and CH$_3$OH spectral windows were fitted simultaneously using an MCMC SUBLIME 1D retrieval to constrain the physical and chemical properties of the coma. The best-fit models are shown in red. Velocity scales are shown in the comet rest frame. For HDO, H$_2$O, CH$_3$OH Band 6 and CH$_3$OH Band 5, the reference frequencies are 241561.550, 183310.087, 241806.524, and 193454.358 MHz, respectively.}
    \label{fig:all_spec1d}
\end{figure*}

We constrain the coma model by fitting the spectra with a Markov chain Monte Carlo (MCMC) approach, following Bayesian-inference techniques similar to those used in exoplanet atmospheric retrievals (for example, refs. \cite{MacDonald:2017MNRAS, Meynardie:2025ApJ}). All detected CH$_3$OH lines in Bands 5 and Band 6 are fitted jointly with the Band 5 H$_2$O and Band 6 HDO spectral regions. This provides a `snapshot’ of the kinetic temperature, expansion velocities and water production rate at the epoch of the ALMA observations, as well as the relative abundances of our three molecules, assuming direct sublimation from the nucleus. We adopt a parent (nucleus) source distribution as a baseline assumption appropriate for observations obtained near perihelion, when outgassing is expected to be comparatively stronger and the coma to expand quasi-isotropically. Consistent with this assumption, the ALMA maps show no significant evidence for spatial heterogeneity, although the beam size prevents definitive constraints. The panels in Figure~2 show the ALMA spectra together with the corresponding best-fit models obtained from a simultaneous fit to the HDO, H$_2$O and CH$_3$OH observations. A detailed description of the radiative transfer modeling and MCMC fitting procedure is provided in Methods.

Our framework enables a robust exploration of the relevant parameter space and allows us to derive posterior probability distributions for the physical properties of the coma and the molecular abundances (Extended Data Table~2 and Extended Data Figure~1). As global model parameters, we derived a coma kinetic temperature of $T_\mathrm{kin}=70.4^{+5.0}_{-4.6}$ K, and a water production rate of $Q(\mathrm{H_2O})=(1.6^{+0.2}_{-0.2})\times10^{29}$ s$^{-1}$. Despite the absence of a formal spectroscopic detection of H$_2$O, our MCMC retrieval is able to constrain both an upper and a lower bound on the H$_2$O production rate indirectly. This is due to the observation of multiple CH$_3$OH lines from different $J_K$ upper-state energy levels (Extended Data Figure~2), namely, the $J_K=5_K-4_K$ band at 242~GHz and the $J_K=4_K-3_K$ band, whose level populations are governed by a balance of collisional and radiative processes \cite{Bockelee-Morvan:1994}. As the population of $J$-states remains closer to LTE for longer than the $K$ states, the combination of both sets of CH$_3$OH lines provides sufficient information to constrain both the gas kinetic temperature and the collision rate. The collision rate, in turn, reveals the density of collision partners, and therefore, the total gas production rate of the coma. To confirm this result, we performed an independent MCMC retrieval, including only the CH$_3$OH spectra, without H$_2$O or HDO, and obtained consistent results within the credible intervals, thus demonstrating that the water production rate is primarily constrained by the CH$_3$OH line excitation. 

It is important to keep in mind the caveats of our approach to measuring $Q(\mathrm{H_2O})$. In particular, our use of an average cross-section for the calculation of state-to-state H$_2$O-CH$_3$OH collisional rates introduces inherent uncertainties into the CH$_3$OH excitation calculation, with associated uncertainties on the $Q(\mathrm{H_2O})$ bounds. Most importantly, this calculation assumes that H$_2$O is the main collisional partner in the coma. Institut de radioastronomie millimétrique (IRAM) 30-m observations by ref. \cite{Biver:2026arXiv} indicate a relatively modest carbon monoxide (CO) production rate near perihelion, suggesting that CO was probably not the dominant collider at the time of our ALMA observations. JWST observations before and after perihelion revealed CO$_2$ as the dominant coma gas \cite{Cordiner:2025ApJ, Belyakov:2026}. However, these measurements were obtained at much larger heliocentric distances, when 3I/ATLAS was not fully inside the H$_2$O iceline, and therefore when H$_2$O sublimation was less efficient \cite{Meech:2004come.book}. In contrast, the Solar Wind ANisotropies (SWAN) camera on the Solar and Heliospheric Observatory (SOHO) shows a dramatic increase in $Q(\mathrm{H_2O})$ close to perihelion relative to pre-perihelion constraints \cite{Combi:2026ApJ}. Extrapolating the JWST Mid-Infrared Instrument (MIRI) post-perihelion water production rates \cite{Belyakov:2026} likewise implies $Q$(H$_2$O)$>$$Q$(CO$_2$) at the order-of-magnitude level near perihelion (Methods and Extended Data Figure~3). Nevertheless, a non-negligible CO$_2$ abundance may still have been present at the time of our observations, consistent with the low coma expansion velocities derived for CH$_3$OH and HDO. This may indicate that molecules heavier than H$_2$O contributed to driving the coma outflow \cite{Biver:2026arXiv}. In light of these uncertainties, we treat the water production rate inferred from CH$_3$OH excitation as an upper limit, $Q(\mathrm{H_2O})<1.6\times10^{29}$ s$^{-1}$. See Methods for a more complete discussion of the caveats and interpretation of this result; further development of this technique for determining water production rates in comets is left to future work.

As an additional conservative scenario, we performed a separate retrieval using only the H$_2$O and HDO spectral data, from which we derived a 99th percentile upper limit of $Q(\mathrm{H_2O})<6.5\times10^{28}$ s$^{-1}$. For the remainder of this paper, we consider both scenarios, with a $Q(\mathrm{H_2O})$ upper limit constrained using the joint HDO+H$_2$O+CH$_3$OH fit, and a more conservative scenario based on the upper-limit constraint derived from the H$_2$O+HDO fit. The resulting best-fit parameters are largely consistent with independent constraints (Extended Data Figure~3) on (1) the water production rate derived from contemporaneous observations with SOHO/SWAN and the Imaging Ultraviolet Spectrograph (IUVS) on board the Mars Atmosphere and Volatile EvolutioN (MAVEN) spacecraft \cite{Combi:2026ApJ}, (2) the CH$_3$OH production rate inferred from pre-perihelion ALMA observations \cite{Roth:2026ApJ}, and (3) the CH$_3$OH production rate measured with the IRAM 30-m telescope close in time to our observations \cite{Biver:2026arXiv}. Further details on the comparison with the independent datasets, are provided in Methods.

\begin{figure*}
    \centering
    \includegraphics[width=1\linewidth]{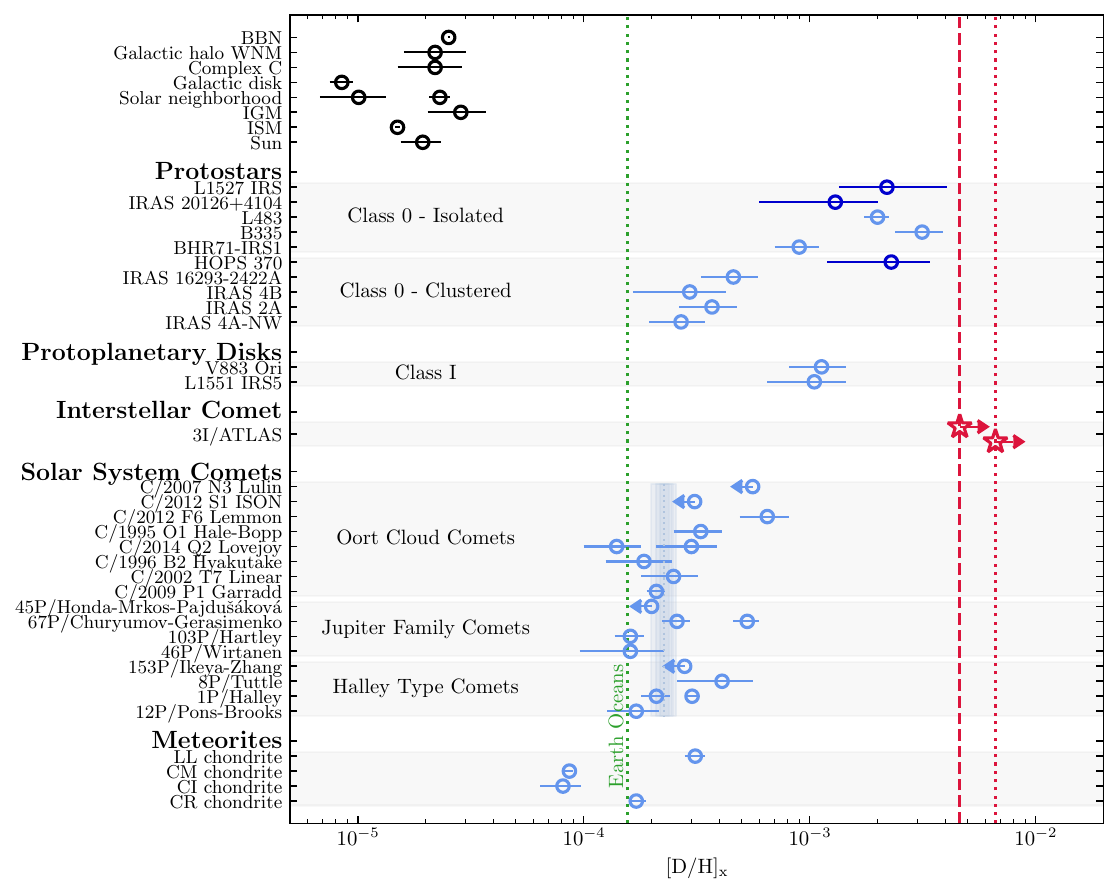}    
    \caption{\textbf{Comparison of D/H values across the Galaxy from atomic or molecular hydrogen, and water D/H values measured in protostars, protoplanetary disks, Solar System comets, meteorites, and 3I/ATLAS}. Bulk atomic/molecular D/H measurements are shown as black points. Gas-phase water D/H measurements in different star- and planet-forming environments are shown in blue, and water D/H measurements in protostars derived from ice (dust-grains) observations are shown in dark blue. Meteorite D/H values are derived from hydrated minerals. Points represent the reported central values, and error bars denote their quoted 1$\sigma$ uncertainties (see Methods for sources). This work constrains the gas-phase water D/H ratio in the interstellar comet 3I/ATLAS (red) from ALMA observations near perihelion. The red star with a dashed vertical line shows the lower limit obtained from the joint HDO+H$_2$O+CH$_3$OH fit, which yields a model-dependent determination of the water production rate. The red star with a dotted vertical line shows the lower limit derived using the upper limit on the water production rate from a fit to the H$_2$O and HDO datasets. BBN: Big Bang nucleosynthesis; WNM: Warm neutral medium; IGM: Intergalactic medium.}
    \label{fig:D2H}
\end{figure*}

From our non-LTE Bayesian retrieval, in the nominal scenario, we derive a lower limit on the HDO/H$_2$O mixing ratio of $x$(HDO)$>9.2\times10^{-3}$. After accounting for the number of H atoms in the water molecule and converting the result to the D/H ratio, we derive a water D/H constraint for the interstellar comet 3I/ATLAS of [D/H]$_{\mathrm{H_2O}} > 4.6\times10^{-3}$. For the conservative scenario, using the 99th percentile upper limit on the water production rate, we derive a lower limit on the D/H ratio of [D/H]$_{\mathrm{H_2O}} > 6.6\times10^{-3}$.  Relative to Earth's ocean value (Vienna Standard Mean Ocean Water), these imply that 3I/ATLAS is enriched in deuterated water by factors of $\gtrsim30$ and $\gtrsim40$, respectively, in the two scenarios. As illustrated in Figure~3, the inferred water D/H for 3I/ATLAS lies at the high end of the observed water D/H distribution, exceeding the error-weighted mean cometary value by factors of $\gtrsim20$ and $\gtrsim30$. Regardless of whether we adopt the nominal or conservative scenario, our constraint on the water D/H ratio in 3I/ATLAS implies a remarkably high D enhancement compared with Solar System comets. Note that the constraint reported in this work corresponds to a single epoch; however, preliminary analysis of ALMA observations obtained close in time suggests no significant day-to-day variability. Even if modest unaccounted-for small-scale variability were present, it would not affect the main finding of this work, namely, that water in 3I/ATLAS is D enriched relative to Solar System comets. This interpretation is supported by independent JWST Near Infrared Spectrograph post-perihelion observations obtained more than a month later, which likewise imply a D enrichment in 3I/ATLAS water \cite{Cordiner:2026arXiv}. 

The strong enrichment of the D/H ratio in water observed in 3I/ATLAS cannot be attributed solely to variations in the bulk atomic/molecular hydrogen composition of the interstellar cloud in which its parent system formed. The primordial atomic D/H ratio was set during the first minutes of the Universe's history, during Big Bang Nucleosynthesis \cite{Epstein:1976Natur, Steigman:2007ARNPS, Mathews:2017IJMPE}. Subsequent Galactic evolution gradually reduces the bulk hydrogen D/H abundance by destroying D in stellar interiors, with the processed material later returned to the interstellar medium (ISM) through stellar winds and supernova explosions \cite{Dvorkin:2016MNRAS, vandeVoort:2018MNRAS}. Additional variations in bulk D/H along different Galactic sightlines may arise if the Milky Way is still accreting relatively pristine, low-metallicity gas with high D content \cite{Wakker:1999Natur, Friedman:2023ApJ}.  The molecular D/H in the ISM provides the initial chemical conditions from which fractionation processes later incorporate deuterium into water in the gas and/or ice phase. The parent star of 3I/ATLAS cannot be reliably identified by backward orbital propagation, in part because we lack sufficiently precise six-dimensional phase-space information for most stars \cite{Bailer-Jones:2018AJ, Hallatt:2020AJ, Guo:2025AJ}. Even if the Galactic location of origin of 3I/ATLAS could be identified, the measured variations in the gas-phase atomic D/H ratio measured along different sightlines \cite{Savage:2007ApJ} are modest (Figure~3). This implies that the water D/H enrichment relative to Solar System comets that we infer for 3I/ATLAS cannot be explained by different initial conditions in the interstellar medium, such as simply placing its birth system in a different region (for example, thick disk) or epoch (for example, 11 Gyr ago) of the Galaxy.

The elevated water D/H ratio in 3I/ATLAS might, in principle, reflect processing after the object was ejected from its natal planetary system. For example, the measured composition could include a layer of interstellar material accreted during passages through cold interstellar clouds \cite{Combi:2026ApJ}. However, the ISM range of bulk atomic or molecular D/H values \cite{Prodanoviv:2010MNRAS} remains well below the enhancements characteristic of water in star- and planet-forming environments, from protostars to Solar System comets and now 3I/ATLAS (Figure~3). It has also been suggested that the volatile composition observed in 3I/ATLAS pre-perihelion could arise from a crust of cosmic-ray-processed material \cite{Maggiolo:2026ApJ}. Although cosmic-ray spallation can produce deuterium \cite{Epstein:1976Natur}, it is unlikely that this mechanism could operate efficiently in the upper layers of ices and refractories exposed during its extended inactive residence in the ISM.

Astronomical observations and detailed models demonstrate that D fractionation may be facilitated within the parental dense core and in the outer regions ($\gtrsim$ 10~au) of planet-forming disks \cite{Ceccarelli:2014prpl.conf, Nomura:2023ASPC}. The central factor required to form water with elevated D/H ratios is cold temperatures (T $<$ 30~K), which is the case for both the natal core and the outer disk.  Other factors are the presence of atomic oxygen and a source of ionization within the bulk of the mass. As a result, the D-rich water ice preserved in 3I/ATLAS and in Solar System comets can be interpreted as either primordial, inherited from the earliest phases of star formation, or as having been reprocessed during the protoplanetary disk phase. In this context, the highly elevated water D/H ratio inferred for 3I/ATLAS relative to Solar System comets points to important differences in the physical and chemical processing of water across the star- and planet-formation sequence in the Solar System versus the 3I/ATLAS host system.

The highest levels of deuteration in water are expected to occur in the prestellar and early protostellar phases, rather than during the initial formation of water within molecular clouds \cite{Furuya:2016A&A}.
In these regions, the temperature of the quiescent gas is influenced by the overall radiation field, which depends on whether the star is born in a cluster (associated with nearby massive stars) or in relative isolation.
In a clustered environment, the radiation field from nearby massive stars heats the high-density filaments \cite{Li:2003ApJ, Kirk:2017ApJ, Hacar:2018A&A} and raises gas temperatures in the filamentary material to 20--30~K, compared with the 10 K more typical of isolated star formation. Because D fractionation is strongly temperature sensitive, these differences should translate into distinct levels of D enrichment \cite{Furuya:2017A&A, Jensen:2021A&A, Jensen:2021A&A2, Bergin:2024come.book....3B}. The Sun most likely formed in a clustered environment, supported by the presence of short-lived radionuclides that may require injection from a nearby supernova \cite{Adams:2010ARA&A, Desch:2024SSRv} and by dynamical signatures such as the high eccentricities of Sedna-like objects and the apparent truncation of the Kuiper Belt, both plausibly produced by an early stellar encounter \cite{Kobayashi:2001Icar, Kenyon:2004Natur, Brasser:2006Icar, Ida:2000ApJ}. Early observations of protostars in star-forming complexes find gas-phase water D/H ratios that are a factor of two to four lower than those measured in isolated Class 0 protostars, but consistent with Solar System comets \cite{Persson:2014A&A, Jensen:2019A&A}. This is interpreted as additional evidence for Solar System formation in a clustered environment. However, recent observations of HDO in the ice phase are challenging this view, with an elevated water D/H ratio measured in clustered protostars \cite{Andreu:2023A&A, Slavicinska:2024A&A, Slavicinska:2025ApJ}. 

Theoretical models suggest that the origin of water D/H enrichment is tied to prestellar stages of water formation, rather than to in situ midplane chemistry in the parent disk \cite{Cleeves:2014Sci}. The first reason is that the ionization level in the bulk disk is reduced due to the exclusion of low energy (molecule ionizing) cosmic rays by energetic stellar winds. Second, oxygen atoms are not found in abundance in outer disk gas, as water is formed in the natal core and remains in the solid state during disk formation and subsequent evolution \cite{Cleeves:2014Sci}. If the water D/H ratio is set mainly during the earliest prestellar phases, the strikingly different water D/H ratio inferred for 3I/ATLAS relative to Solar System comets points to a markedly different prestellar environment, most plausibly colder conditions that favor stronger fractionation.

An alternative to a primordial origin of the water D/H enhancement is processing within the protoplanetary disk. In this picture, relatively pristine (D/H enriched) ices that originated in cold regions are blended with water that has been thermally processed or isotopically re-equilibrated at higher temperatures, thereby diluting the original prestellar D/H ratio \cite{Zannese:2024NatAs}. Radial mixing driven by angular-momentum transport can carry warm, inner-disk gas outwards and mix it with colder outer-disk material \cite{Yang:2013Icar}. The relevance of this effect remains uncertain, especially if late-stage infall persists \cite{Huang:2023ApJ, Winter:2024A&A}, which could continuously supply D-enriched material and limit any substantial reset. Another complication is that the maximum D/H ratio achievable for ice water on presolar grains is still unknown. If the water in both 3I/ATLAS and Solar System comets began with the same prestellar D/H ratio, then the difference we infer would point to markedly different evolutionary histories during the protoplanetary disk stage. In this disk-reprocessing scenario, an early ejection of 3I/ATLAS from its host system could be consistent with a higher retained D/H ratio than the Solar System comet population observed today.

Solar System comets formed at broadly similar distances within the giant-planet region \cite{Brasser:2013Icar, Kaib:2024come.book}, probably within or in proximity to the CO and CO$_2$ snowlines \cite{AHearn:2012ApJ}. If comet formation location within a protoplanetary disk translates into different levels of D enrichment, then the elevated D/H ratio inferred for 3I/ATLAS could indicate a formation location beyond the CO$_2$ snowline in its parent disk. This interpretation would be consistent with the CO$_2$ enrichment reported from JWST and SPHEREx \cite{Cordiner:2025ApJ, Lisse:2025RNAAS}. Although early theoretical work proposed that the D/H ratio increases monotonically with formation distance in the protosolar nebula \cite{Drouart:1999Icar, Hersant:2001ApJ, Kavelaars:2011ApJ}, such a gradient may be erased by radial mixing \cite{Yang:2013Icar}. Indeed, the absence of a strong protoplanetary water D/H gradient is consistent with the relatively modest scatter in water D/H across different comet populations. In any case, a formation location in the outer disk for 3I/ATLAS could also help explain its delivery as an interstellar object, because planetesimals at large radii from their protostar are expected to be more readily ejected by planet scattering and stellar flybys than bodies formed in the inner disk \cite{Raymond:2010ApJ, Hands:2019MNRAS, Fitzsimmons:2024come.book}.

Other factors could potentially influence the water D/H ratio during the main phase of water-ice formation in the prestellar cloud. These could be related to the collapse timescales, the overall exposure of the ultraviolet radiation field, and/or changes in the cosmic-ray ionization rate \cite{Nomura:2023ASPC}. However, given the strong temperature dependence of the initiating D fractionation reactions, it is highly likely that the natal environment of 3I/ATLAS was colder than the solar nebular cloud at the time of the Sun's birth. Even if disk reprocessing plays an important role, the elevated water D/H in 3I/ATLAS relative to Solar System comets would then imply a different protoplanetary disk history, with less efficient thermal and isotopic resetting. In either case, whether the D/H ratio was inherited from the prestellar phase or modified within the disk, the physical and chemical conditions that shaped 3I/ATLAS were remarkably different from those that shaped Solar System comets.  3I/ATLAS and its D/H ratio not only extend D/H constraints beyond the Solar System, supporting the presence of this isotopic signature in water in other planetary systems, but also highlight the diversity of outcomes for planetary birth, and that the composition and history of solids in other solar systems can be different from our own.

\section*{Methods}\label{sec2}

\subsection*{ALMA Data}\label{subsec1}

The dataset presented in this work was obtained through the ALMA DDT program 2025.A.00002.T (principal investigator T.P.-C.) on 4 November 2025 using the ACA. The main molecular targets of the program were H$_2$O ($J_{K_{a},K_{c}}$ = 3$_{1,3}$ -- 2$_{2,0}$ at $\nu$ = 183.310\,GHz) in ALMA Band 5 and HDO ($J_{K_{a},K_{c}}$ = 2$_{1,1}$ -- 2$_{1,2}$ at $\nu$ = 241.561\,GHz) in ALMA Band 6, together with multiple CH$_3$OH transitions in the 193-GHz and 241-GHz windows. Our program had four spectral windows in each band. For the spectral setting of Band 5, two broad continuum windows with a bandwidth of 1.875\,GHz were centered at 183.300 and 193.500\,GHz. Our target H$_2$O and CH$_3$OH lines were sampled at a resolution of 122.07\,kHz with spectral windows centered at 183.310 and 193.480\,GHz, respectively. In Band 6 a similar setting is used, with broad continuum windows of 1.875\,GHz centered at 226.995 and 228.745\,GHz, while the target HDO and CH$_3$OH lines are observed at a resolution of 244.14\,kHz in overlapping windows centered at 241.562 and 241.970\,GHz.

The specific data corresponding to each date of observation can be found in Extended Data Table~1. In the execution block of Band 5, 20\,min were performed on source, while in Band 6 it was 50\,min on source. The quasar J1337-1257 was used as calibrator for all observations, maintaining a steady flux value throughout the observing period. The mean precipitable water vapor at zenith on 4 November ranged from 0.386-0.405 mm. 

The location of the comet was predicted based on the latest available ephemeris solution from the JPL Horizons database before the observations (ref. JPL \#27). For all datasets, the nucleus of 3I/ATLAS appeared offset by 5$''$ to the east and 0.8$''$ south from the expected ephemeris position. This offset was an early indication of the presence of non-gravitational forces perturbing the trajectory of 3I/ATLAS. All of the datasets were reduced using the ALMA pipeline version 2025.1.0.35 and CASA version 6.6.5. Imaging and deconvolution were conducted using CASA task \texttt{tclean} and natural weighing to maximize signal to noise. A 8$''$-diameter circular mask was used and a stop flux threshold of 2 times the channel RMS and 1 times the channel RMS for Band 6 and Band 5 datasets, respectively.  In the final image cubes, HDO and all CH$_3$OH lines were confidently detected at similar strengths, while H$_2$O was not detected.

To extract the spectra, as no continuum flux was detected, we identified the pixel corresponding to the 3I/ATLAS nucleus using line emission maps, as described below. For every spatial pixel, we estimated the noise level from the spectrum as $\sigma_{\mathrm{rms}} = 1.4826\,\sigma_{\mathrm{MAD}}$, where $\sigma_{\mathrm{MAD}}$ is the median absolute deviation across channels. This robust estimator is appropriate because the majority of channels in the spectral windows are line emission free. We then selected all channels with emission exceeding $3\sigma$ and collapsed the cube along the spectral axis, thereby constructing a two-dimensional (2D) image. A symmetric 2D Gaussian with a fixed width matching the synthesized beam was fitted to this image to locate the centroid of the emission, which we adopted as the nucleus position. This procedure was applied to the Band 6 data to determine the extraction pixel for both the HDO and CH$_3$OH spectra, and separately to the Band 5 spectral window with CH$_3$OH. We extracted the H$_2$O spectrum at the position determined from the Band 5 CH$_3$OH window, which was observed simultaneously. Note that although we use single-pixel spectra, each pixel is in Jy beam$^{-1}$ and thus already corresponds to flux integrated over the synthesized ACA beam.

\subsection*{Line Emission Modeling}\label{subsec1}

We modeled the line emission using the SUBLIME radiative transfer code \cite{Cordiner:2022ApJ}. SUBLIME considers the outflow dynamics of an expanding cometary coma and the detailed balance between collisional and radiative processes to predict the rotational emission of the gas. We analyzed the dataset reported in this work with the 1D version of SUBLIME, which assumes an isotropically expanding gas with constant outflow velocity, kinetic temperature and production rate \cite{Haser:1957BSRSL, Haser:2020PSJ, Cordiner:2024PSJ, Coulson:2026MNRAS, Roth:2026ApJ}. We adopted the 1D version instead of the more detailed 3D version because it has already been shown for 3I/ATLAS that 1D and 3D SUBLIME modeling provide mutually consistent results \cite{Roth:2026ApJ}. The applicability of 1D models is a consequence of the low molecular abundances in a cometary atmosphere. The optically thin assumption is supported by our data, given the relative symmetry of the spectral line profiles (Figure~2 and Extended Data Figure~2), suggesting that a 1D model, widely adopted for cometary rotational lines, is a reasonable approximation. The modeled brightness temperatures are low compared with the inferred rotational and kinetic temperatures, and the beam-averaged optical depths computed from the best-fit models indicate $\tau\ll1$. Full 3D modeling of the complete 10-day dataset from ALMA DDT program 2025.A.00002.T, together with the ALMA DDT program 2025.A.00004.S (principal investigator M.A.C.), is deferred to future work.

Our non-LTE calculations use the same collisional and radiative configuration as in ref. \cite{Cordiner:2025NatAs}. Energy levels and radiative transition data for HDO, H$_2$O, and CH$_3$OH are taken from the Leiden Atomic and Molecular Database LAMBDA \cite{Schoier:2005A&A, vanderTak:2020Atoms}. For H$_2$O, we use state-to-state p-H$_2$O–H$_2$O collisional rate coefficients from ref. \cite{Mandal:2024A&A}. The HDO and CH$_3$OH lines are modeled using the thermalization approximation \cite{Crovisier:1987A&AS, Biver:1999AJ, Bockelee-Morvan:2012A&A}, with an average H$_2$O collision cross-section of $5\times10^{-14}$ cm$^{2}$. Collisions with electrons are included via the Born approximation \cite{Itikawa:1972JPSJ}, adopting an electron density scaling $x{_\mathrm{ne}}=0.2$ (refs. \cite{Hartogh:2010A&A, Biver:2019A&A}). Solar pumping rates are obtained from the Planetary Spectrum Generator (PSG) \cite{Villanueva:2018JQSRT} and rescaled to the heliocentric distance of 3I/ATLAS at the time of the ALMA observations. Photodissociation rates appropriate for the active Sun were taken from ref. \cite{Huebner:2015P&SS}. For all simulations, we use the escape probability approximation \cite{Bockelee-Morvan:1987A&A}, which is particular important for treating radiation trapping in water \cite{Bensch:2004ApJ, Zakharov:2007A&A}. A flux loss factor due to the ACA configuration is estimated to be 0.62. This is calculated by using the \texttt{simobserve} CASA task to create a mock observation using the same antenna positions and integration time as the real data. The mock spectral cube is then imaged with \texttt{tclean} using the same parameters as for the 4 November data and compared with the input file to calculate the flux loss value.

% A water photodissociation rate of 2.2 $\times$ 10$^{-5}$ s$^{-1}$ appropriate for the active Sun was taken from \cite{Huebner:2015P&SS}.

\subsection*{Bayesian-inference fitting}\label{subsec1}

Non-LTE effects produce strong, nonlinear couplings and degeneracies among the physical parameters that describe a cometary atmosphere. To characterize these reliably, we adopted an MCMC approach to fit our data. Specifically, we used the affine invariant ensemble sampler \cite{Goodman:2010CAMCS} as implemented in the \texttt{emcee} package \cite{Foreman-Mackey:2013PASP}. Within this Bayesian framework, our chosen priors are combined with the likelihood of the data to obtain posterior probability distributions for each parameter. The ensemble sampler efficiently explores the full parameter space.

We construct the total log-likelihood under the assumption of Gaussian errors. The total likelihood is written as the sum of the contributions from each molecule or dataset $X$, where $X$ can be H$_2$O, HDO, CH$_3$OH Band 5, or CH$_3$OH Band 6. For fitting purposes we treat the CH$_3$OH emission in the two Bands as separate datasets, but we impose the same physical and chemical parameters for both. The general form of the likelihood is:

\begin{equation}\label{eq:likelihood}
\begin{aligned}
&\ln \,\mathcal{L} (\Theta) = -\frac{1}{2} \sum_{X}^{N_{\text{mol.}}} \chi^2_X(\Theta) \\
&= -\frac{1}{2} \sum_{X}^{N_{\text{mol.}}} \sum_{v}^{N_{\text{chan.}}} \left( \frac{F_X(v) - M_X(v \mid \Theta)}{\sigma_X} \right)^2,
\end{aligned}
\end{equation}

where $F_X(v)$ denotes the observed ALMA spectrum of dataset $X$ as a function of velocity, $M_X(v \mid \Theta)$ is the corresponding SUBLIME model spectrum for a parameter vector $\Theta$, and $\sigma_X$ is the MAD-derived noise adopted for dataset $X$. In the MCMC sampling of $\ln \mathcal{L}(\Theta)$ we typically used a number of walkers equal to three times the number of free parameters. We assumed uninformative uniform priors within physically motivated bounds (Extended Data Table~2); for parameters spanning orders of magnitude (that is, molecular production rates) we used log-uniform priors. Walkers were initialized by drawing starting positions uniformly across the allowed parameter ranges.     

\subsection*{Cometary atmosphere retrieval}

To derive a self-consistent physical and chemical model of the 3I/ATLAS coma, we take advantage of the near-simultaneity of our HDO, H$_2$O, and CH$_3$OH observations. The Band 6 dataset was obtained on 4 November starting at 13:00 UTC, and the Band 5 observations, also on 4 November, ended at 15:31 UTC, with an interband separation of $\sim10$ min. The total time span of this dataset represents only a small fraction of the comet’s rotation period of 16.16$\pm$0.01 hours \cite{Santana-Ros:2025A&A}, so rotational modulation and large-scale evolution of the coma are expected to be minimal. We note, however, that CH$_3$OH emission in Solar System comets can vary periodically on minute-to-hour timescales, probably reflecting changing outgassing geometries of active regions as the comet nucleus rotates \cite{Roth:2021PSJ, Cordiner:2023ApJ}. Given the relatively small ratio between the observation spacing and the rotation period, we assume that the combined Band 5 and Band 6 datasets sample the same underlying state of the coma.

We performed a joint MCMC retrieval that simultaneously fits the H$_2$O window, together with the HDO emission and the CH$_3$OH emission in Bands 5 and 6. We assume a parent model for each molecule; therefore, the parameter vector $\Theta$ includes a total of eight free parameters. The global parameters are the coma kinetic temperature $T_\mathrm{kin}$ and the H$_2$O production rate $Q({\rm H_2O})$. Molecule-specific parameters are the production rate, expansion velocity and Doppler shift of both CH$_3$OH and HDO (that is, $Q({\rm CH_3OH})$, $v_{\rm exp,\,CH_3OH}$, $\Delta v_\mathrm{CH_3OH}$, $Q({\rm HDO})$, $v_{\rm exp, \, HDO}$, $\Delta v_\mathrm{HDO}$). The prior ranges adopted for each parameter in this fit are summarized in Extended Data Table~2. As the H$_2$O emission line was not detected, we fix the kinematics of H$_2$O (that is, $v_\mathrm{exp}$ and $\Delta v$) to those of its isotopologue HDO. For the water ortho-to-para ratio (OPR), we assume statistical equilibrium \cite{Hama:2016Sci}, motivated by the fact that most cometary measurements are consistent with OPR = 3 (refs. \cite{Bonev:2007ApJ, Villanueva:2011Icar, Bonev:2013Icar, Cheng:2022A&A, Cordiner:2025NatAs}), and by OPR constraints from modeling of ortho- and para-H$_2$O rovibrational lines in JWST post-perihelion observations (N. Roth, personal communication, and ref. \cite{Belyakov:2026}). 

The best-fit posterior values are summarized in Extended Data Table~2, and the joint posterior distributions together with their marginal posteriors are shown in Extended Data Figure~1.

\subsection*{CH$_3$OH emission}

In Extended Data Figure~2, we compare the CH$_3$OH ALMA spectra with the corresponding best-fit model. The ensemble of CH$_3$OH lines included in the retrieval provides strong leverage on both the coma kinetic temperature and its expansion velocity \cite{BockeleeMorvan:2004come.book, Biver:2021A&A}. 

Despite the overall good agreement between the model and the data, the high-resolution Band 5 spectra in Extended Data Figure~2 show a clear asymmetry on the blue side of the two strongest lines that cannot be fully captured by our assumed spherically symmetric SUBLIME 1D model. The Band 5 CH$_3$OH model underpredicts the observed intensities by $\sim20\%$, which could reflect unaccounted non-LTE effects, time variability between the two observing windows and uncertainties in ALMA's absolute flux calibration. A more detailed 3D treatment of this dataset, and of our additional ALMA observations, will be presented in future work. 

To validate our CH$_3$OH modeling, we compare our derived CH$_3$OH production rate with independent measurements. Our HDO+H$_2$O+CH$_3$OH retrieval yields $Q$(CH$_3$OH)$=(4.1\pm0.1)\times10^{27}$ s$^{-1}$. The ALMA monitoring results of ref. \cite{Roth:2026ApJ} were obtained pre-perihelion and thus are not directly comparable; however, our value is consistent with an extrapolation of the increasing trend reported in that study (Extended Data Figure~3). In addition, IRAM 30-m observations obtained closer in time (1-3 November) yield $Q$(CH$_3$OH)$=(3.38\pm0.05)\times10^{27}$ s$^{-1}$ (ref. \cite{Biver:2026arXiv}), in good agreement with our inferred CH$_3$OH production rate.

The rotational temperature ($T_{\rm rot}$) of the observed CH$_3$OH gas was also compared with the model results through a rotational diagram analysis as is commonly done in the literature \cite{Bockelee-Morvan:1994, Roth:2021PSJ}. The rotational diagram is constructed by assuming that the population distribution of the levels sampled by CH$_3$OH follow a Boltzmann distribution characterized by a single temperature, $T_{\rm rot}$. In this case the column
density of molecules in the upper level of each transition $N_{\rm u}$ can be related to the integrated flux density $S_{\nu}\Delta v$ via:

\begin{equation}
    N_{\rm u} = \frac{4\pi S_{\nu}\Delta v}{A_{\rm ul}\Omega h c}.
\end{equation}

Where $\Omega$ is the subtended angle by the emission, which we assume fills the full beam in all cases, $A_{\rm{ul}}$ is the Einstein A coefficient, $h$ is the Planck constant and $c$ is the speed of light. This can then be related to the total column density of the
molecular emission through the Boltzmann distribution

\begin{equation}
    \frac{N_{\rm u}}{g_{\rm u}} = \frac{N_{\rm T}}{Q(T_{\rm rot})} e^{-E_{\rm u}/k_{\rm B}T_{\rm rot}}
\end{equation}

$E_{\rm u}$ and $g_{\rm u}$ are the upper-level energy and degeneracy, respectively. $N_{\mathrm T}$ is the total column density, and $k_{\mathrm B}$ is the Boltzmann constant. $Q(T_{\rm rot})$ is the partition function at the rotation temperature, obtained by interpolating from the tabulated values in the Cologne Database for Molecular Spectroscopy (CDMS) \cite{Muller:CDMS}. Taking the logarithm of the previous equation, it is possible to derive both the total column density and the rotational temperature. This approach assumes optically thin emission, which is appropriate for the CH$_3$OH ALMA observations.

Supplementary Figure~1 shows the results for the rotational diagrams constructed by accounting for all CH$_3$OH transitions observed in each band ($J = 4 - 3$, 193 GHz and $J  = 5 - 4$, 242 GHz). The data has a rotational temperature of $\sim$38\,K in both bands and even though the joint HDO+H$_2$O+CH$_3$OH model was computed with a kinetic temperature of 70.4\,K, it traces a rotational temperature of $\sim$40\,K. Within uncertainties, the rotational temperature of the model and data are consistent. The lower value with respect to the kinetic temperature likely indicates that we are sampling CH$_3$OH molecules that have left the collisional region and are not in LTE.

\subsection*{$Q$(H$_2$O) constraint}

Our MCMC SUBLIME 1D retrieval from the joint HDO+H$_2$O+CH$_3$OH fit yields a water production rate of $Q({\rm H_2O})=\left(1.6^{+0.2}_{-0.2}\right)\times10^{29}\ {\rm s^{-1}}$, consistent with independent constraints near perihelion. The earliest post-perihelion SOHO/SWAN measurements (November 6 and 8) give $3.17\times10^{29}$ and $2.31\times10^{29}$ s$^{-1}$, respectively \cite{Combi:2026ApJ}. Although these observations are offset from ours by 2-4 days and may sample different rotational phases, their agreement within a factor of $\sim2$ is indicative of the consistency of our inferred $Q({\rm H_2O})$. Because SOHO/SWAN uses a much larger field of view than most other instruments, its inferred production rates can differ from smaller-aperture measurements when dust-grain (extended-source) contributions are present \cite{Ejeta:2024AJ}; we discuss extended sources explicitly in the next subsection. The SOHO/SWAN production rates remain approximately steady from 9 days (their first measurement) to 22 days after perihelion (Extended Data Figure~3), with a median of $2.35\times10^{29}$ s$^{-1}$ and a standard deviation of $4.07\times10^{28}$ s$^{-1}$. However, note that H Lyman-$\alpha$ observations are not sensitive to short-term variations because of photochemical timescales. Reference \cite{Combi:2026ApJ} further reports preliminary MAVEN/IUVS \cite{McClintock:2015SSRv} results, indicating water production rates of $\sim(2.2-3.2)\times10^{29}$ s$^{-1}$ over a similar time span, extending to 6 days before perihelion. Taken together, two independent instruments, MAVEN/IUVS and SOHO/SWAN, which infer $Q({\rm H_2O})$ indirectly from two independent tracers, H Lyman-$\alpha$ and OH emission \cite{Makinen:2005Icar, Crismani:2015GeoRL, Combi:2019Icar, Mayyasi:2020AJ}, are in agreement with our derived water production rate around perihelion. 

The ability to constrain the water production rate in the absence of a formally significant H$_2$O detection is a consequence of our self-consistent MCMC treatment of the CH$_3$OH and water datasets. Two complementary pieces of information drive this constraint. The Band 5 H$_2$O spectral window sets an upper limit through the absence of detectable line emission. Meanwhile, the rich CH$_3$OH spectrum provides an indirect constraint on $Q({\rm H_2O})$ assuming that water is the dominant collisional partner that governs the non-LTE excitation of CH$_3$OH in the coma.
 
Water is a key driver of the observed CH$_3$OH rotational spectrum. As our retrieval performs a statistical global optimization, it can explore the allowed parameter space (Extended Data Table~2) exhaustively (but efficiently) and identify the value of $Q({\rm H_2O})$ that, in combination with the other model parameters, best reproduces the full set of CH$_3$OH lines observed with ALMA. The resulting MCMC chains converge to stable solutions (Supplementary Figure~2), indicating that the joint constraints are well behaved. The water production rate inferred in this way is therefore model dependent, in the sense that it is derived indirectly through its role in CH$_3$OH excitation. Other indirect ways to determine water production include photodissociation tracers such as H Lyman-$\alpha$ or OH (A$^2\Sigma^+$-X$^2\Pi$) \cite{Schleicher:1988ApJ, Makinen:2005Icar, Bodewits:2019ApJ}. The relative intensity of slightly asymmetric rotors, such as CH$_3$OH and formaldehyde (H$_2$CO), are highly sensitive to temperature and density \cite{Giannetti:2025A&A}. In molecular clouds, for example, it is standard to use these two molecules to probe the kinetic temperature and H$_2$ spatial density, leveraging the fact that H$_2$ is the dominant constituent (and collider) in star-forming regions \cite{Mangum:1993ApJS, Leurini:2004A&A}.   

To illustrate that CH$_3$OH indirectly constrains the water abundance of 3I/ATLAS through collisional excitation, we carried out an additional retrieval that excludes the H$_2$O Band 5 and HDO Band 6 data. Specifically, we reran the MCMC SUBLIME 1D analysis using the same global and CH$_3$OH-specific parameters, along with the same prior ranges as in the fiducial fit, but fitting only the CH$_3$OH data in Band 5 and Band 6 (Supplementary Table~1 and Supplementary Figure~3). The resulting best-fit parameters are consistent within uncertainties to those listed in Extended Data Table~2 for the HDO+H$_2$O+CH$_3$OH fit. A comparison of the $Q({\rm H_2O})$ posterior distribution is shown in Supplementary Figure~4. The agreement between the best-fit value and the full posterior shows that the constraint on $Q({\rm H_2O})$ is driven primarily by the CH$_3$OH lines: by exploring the parameter space, the retrieval identifies the non-LTE excitation conditions that best reproduce the ensemble of CH$_3$OH rotational transitions in our quasi-simultaneous Band 5 and Band 6 observations.  

The first caveat regarding our indirect determination of water is our approximate treatment of CH$_3$OH collisions. As there are no published state-to-state H$_2$O-CH$_3$OH collisional rates, SUBLIME 1D calculates them using the thermalization approximation \cite{Crovisier:1987A&AS, Bockelee-Morvan:2012A&A}. We adopt the standard average collisional cross-section of $5\times10^{-14}$ cm$^{-2}$, which, although widely used, remains poorly constrained \cite{Biver:1999AJ}. Another approximation is the treatment of collisions with electrons through the Born approximation \cite{Itikawa:1972JPSJ}, which can be particularly important in the outer coma when collisions between neutrals are infrequent \cite{Biver:2019A&A}. Similarly, we use the widely adopted electron density scaling factor of $x_{\rm ne}=0.2$ (refs. \cite{Zakharov:2007A&A, Hartogh:2010A&A, Val-Borro:2010A&A, Bockelee-Morvan:2012A&A}), which is also uncertain \cite{BockeleeMorvan:2004come.book}. Note that HDO likewise lacks published state-to-state H$_2$O-HDO collisional rates and is modeled under the same assumptions, so any systematic correction affecting one molecule will likely affect the other as they see the same colliders. Additional theoretical and experimental work is needed to improve the treatment of collisional excitation among these species.

The second caveat of our indirect determination of water is the assumption that H$_2$O is the main collisional partner of CH$_3$OH. Water is generally the principal component of the nuclear ices of Solar System comets, followed by CO$_2$, CO, and CH$_3$OH (refs. \cite{Cochran:2015SSRv, Bockelee-Morvan:2017RSPTA}). However, this is not always the case, and the abundance measured in a cometary coma depends not only on the absolute inventory of each molecular species in the nucleus, but also on the location of the observations relative to the different icelines \cite{Ootsubo:2012ApJ, HarringtonPinto:2022PSJ}. For example, the previous interstellar comet, 2I/Borisov, was found to exhibit a higher abundance of CO than water \cite{Cordiner:2020NatAs, Bodewits:2020NatAs}. For 3I/ATLAS, CO observations with the IRAM 30-m telescope obtained between 1 and 3 November yielded $Q$(CO)$=(6.8\pm1.1)\times10^{27}$ s$^{-1}$ (ref. \cite{Biver:2026arXiv}), indicating that CO is not an important collider at the time of our observations ($Q$(H$_2$O)$\sim10^{29}$ s$^{-1}$).

JWST and SPHEREx observations obtained early on the inbound leg of 3I/ATLAS, at $r_H\sim3$ au, indicated that CO$_2$ dominated the volatile outgassing, with CO$_2$/H$_2$O $\sim7$ (refs. \cite{Cordiner:2025ApJ, Lisse:2025RNAAS, Lisse:2026ApJ}). While no CO$_2$ measurements are available contemporaneous with our near-perihelion ALMA data, the earliest post-perihelion SPHEREx observations at $r_H\sim2.1$ au suggest a strong increase in water activity: $Q$(H$_2$O) was larger by a factor of $\sim40$ relative to the pre-perihelion constraint, whereas $Q$(CO$_2$) increased by only $\sim2$ (ref. \cite{Lisse:2026RNAAS}). In addition, JWST/MIRI post-perihelion monitoring between $r_1=2.2$ au and $r_2=2.54$ au found that $Q$(H$_2$O) decreased from $Q_1$(H$_2$O)$=3.78\times10^{27}$ s$^{-1}$ to $Q_2$(H$_2$O)$=1.05\times10^{27}$ s$^{-1}$, whereas $Q$(CO$_2$) declined more gradually, from $Q_1$(CO$_2$)$=8.70\times10^{27}$ s$^{-1}$ to $Q_2$(CO$_2$)$=5.42\times10^{27}$ s$^{-1}$ (ref. \cite{Belyakov:2026}). If we approximate each species as following a linear relation in $\log Q$ versus $\log r$, we can estimate the heliocentric distance $r_0$ at which CO$_2$ and H$_2$O contribute equally to the gas production, $Q_0$(H$_2$O)$=Q_0$(CO$_2$):

\begin{equation}
\begin{aligned}
    \log & \, r_0=\log r_1\\
& + \log\!\left(\frac{r_2}{r_1}\right)\,
\frac{\log\!\left(\frac{Q_1(\mathrm{CO_2})}{Q_1(\mathrm{H_2O})}\right)}
{\log\!\left(\frac{Q_2(\mathrm{H_2O})\,Q_1(\mathrm{CO_2})}{Q_1(\mathrm{H_2O})\,Q_2(\mathrm{CO_2})}\right)}.
\end{aligned}
\end{equation}

Under this approximation, H$_2$O dominates over CO$_2$ after perihelion, and CO$_2$ does not overtake H$_2$O until $r_H\sim1.9$ au. This order-of-magnitude estimate is not intended to provide a realistic determination of the CO$_2$ and H$_2$O production rates near perihelion, but rather to provide a sense of whether H$_2$O is likely the dominant collider for the ALMA epoch ($r_H=1.37$ au). Consistent with H$_2$O becoming the dominant coma gas near perihelion, SOHO/SWAN measurements show more than an order-of-magnitude rise in $Q$(H$_2$O) between $r_H\sim2.1$ au and perihelion \cite{Combi:2026ApJ} (Extended Data Figure~3). Nevertheless, even if H$_2$O is the dominant collisional partner, contributions from CO and CO$_2$ to the collisional excitation budget may not be negligible. The comparatively low expansion velocities inferred for CH$_3$OH and HDO relative to the canonical scaling $v=0.85\,r_H^{-0.5}$ could indicate an outflow influenced by heavier molecules such as CO$_2$ (ref. \cite{Biver:2026arXiv}). If so, the production rate inferred under an H$_2$O-dominated coma assumption may more closely trace, for example, the combined H$_2$O+CO$_2$ gas production rate rather than H$_2$O alone. With the previous caveats in mind, we treat our constraint of the H$_2$O production rate from the CH$_3$OH excitation as an upper limit,  $Q({\rm H_2O})<1.6\times10^{29}\ {\rm s^{-1}}$.

In addition, to present a more conservative scenario, we quantify what our ALMA Band 5 H$_2$O window can constrain on its own by running an additional retrieval using only the H$_2$O and HDO datasets, excluding both CH$_3$OH Bands. We adopted the $T_\mathrm{kin}$ value inferred from the HDO+H$_2$O+CH$_3$OH fit. To propagate the uncertainties from the nominal fit into the H$_2$O+HDO analysis, we treated $T_\mathrm{kin}$ as a free parameter and applied a Gaussian prior with mean and standard deviation set by the corresponding posterior distribution from the HDO+H$_2$O+CH$_3$OH retrieval (Supplementary Table~1 and Supplementary Figure~3). The H$_2$O+HDO run yields a 99th-percentile upper limit of $Q({\rm H_2O})<6.5\times10^{28}$ s$^{-1}$. This limit is lower than the water production rate inferred from the joint HDO+H$_2$O+CH$_3$OH and CH$_3$OH analysis (Supplementary Figure~4).

\subsection*{Extended source contributions}

Even though these observations were obtained with the ACA, which maximizes sensitivity to large angular scales for an ALMA map given the comet geometry, the interferometric response still imposes a finite effective aperture set by the synthesized beam. We estimate an aperture radius by approximating the beam as a 2D Gaussian with an effective full width at half maximum (FWHM) defined as $\mathrm{BMEAN}=\sqrt{\mathrm{BMAJ}\times\mathrm{BMIN}}$, where BMAJ and BMIN are the FWHM of the major and minor axes of the restoring beam, respectively. We then adopt an effective radius of $3\sigma$ (enclosing 99.7$\%$ of the Gaussian), with $\sigma=\mathrm{FWHM}/2.3548$. For the Band 6 HDO and CH$_3$OH data, $\mathrm{BMEAN}=5.26''$; for Band 5 CH$_3$OH, $\mathrm{BMEAN}=7.71''$; and for Band 5 H$_2$O, $\mathrm{BMEAN}=8.10''$. At the comet’s geocentric distance, these correspond to effective radii of approximately 11000 km, 16000 km, and 17000 km. Our forward model accounts for the finite synthesized beam, so the reported $Q$ values correspond to total production rates under the assumption of nucleus-driven sublimation, not the production contained within the beam. Howevever, our apertures remain smaller than the characteristic spatial extent of a cometary coma, so any extended production component on larger scales would be filtered out or strongly down-weighted in our extracted spectra. The moment-0 maps for HDO and CH$_3$OH (Figure~1) are consistent with emission mostly centered on the nucleus. Independent SUBLIME 3D modeling performed directly in the visibility domain likewise finds that the HDO data is consistent with direct sublimation from the nucleus, while CH$_3$OH probably includes contributions from distributed sources (M.A.C. et al., manuscript in preparation), consistent with the parent scale length of $>258$ km determined by ref. \cite{Roth:2026ApJ}.

Because the H$_2$O line is not detected in our ALMA Band 5 spectrum, the water dataset by itself provides no leverage on the presence of an extended water source. An estimate of the active fraction using standard methods \cite{Cowan:1979M&P, AHearn:1995Icar}, based on our derived water production rate, indicates that 3I/ATLAS is a hyperactive comet with a substantial contribution from icy grains. Independent indications of aperture-dependent production rates pre-perihelion, reported from Swift and JWST, are consistent with this picture \cite{Cordiner:2025ApJ, Xing:2025ApJ}. However, our measurements were obtained close to perihelion, when photodissociation scale lengths are expected to be smallest. With the usual scaling $\gamma \propto r_H^2$, the characteristic photodissociation scale length during the JWST pre-perihelion epoch ($r_H=3.32$ au) would be about 6-times larger than during our ALMA epoch ($r_H=1.37$ au). For dust grains, the approximate expression for their scale length, $l_p=250\times a^{0.5}\times(Q(\mathrm{H_2O})/10^{29}\,\mathrm{s^{-1}})^{0.5}$ km, determined by ref. \cite{Bockelee-Morvan:2001Sci} and consistent with detailed calculations \cite{Beer:2006Icar}, where $a$ is the radius of the icy fragments in centimeters, suggests that even for a 1-m fragment, the scale length $l_p\sim4000$ km is within our beam. As a result, even if an extended grain-driven component is present, its impact on our near-perihelion determination of $Q({\rm H_2O})$ should be reduced compared with observations obtained at larger heliocentric distances because some fraction of the H$_2$O released from grains is liberated before leaving our beam. Finally, Solar System comets show an inverse correlation between active fraction and water D/H (ref. \cite{Lis:2019A&A}). If that empirical trend extends to interstellar comets, then epochs when 3I/ATLAS is less hyperactive could exhibit even higher water D/H values, which would be compatible with our D/H lower-limit scenario.

\subsection*{Compilation of D/H values}

The reference D/H values in Figure~3 are compiled as follows. For Big Bang nucleosynthesis (BBN), we use the primordial atomic D/H ratio measured in extremely metal-poor absorption systems along quasar sightlines, D/H $=(2.527\pm0.030)\times10^{-5}$ (ref. \cite{Cooke:2018ApJ}). For the local ISM, we adopt the mean gas-phase atomic D/H in the Local Bubble, D/H $=(1.56\pm0.04)\times10^{-5}$ (ref. \cite{Linsky:2006ApJ}). Atomic D/H estimates for other Galactic environments (Galactic halo warm neutral medium, high-velocity cloud Complex C, Galactic disk, Solar neighborhood, and the intergalactic medium) are taken from the summary in ref. \cite{Savage:2007ApJ}. For the protosolar value, we adopt D/H $=(1.94\pm0.39)\times 10^{-5}$ inferred from Jupiter's atmosphere \cite{Lodders:2003ApJ}. Water D/H measurements in Class 0 protostars (gas and ice) are taken from refs. \cite{Persson:2014A&A, Jensen:2019A&A, Jensen:2021A&A2, Slavicinska:2024A&A, Slavicinska:2025ApJ}, and Class I gas-phase water measurements from refs. \cite{Tobin:2023Natur, Andreu:2023A&A}. Meteorite D/H values (from hydrated minerals in chondrites) are adopted from ref. \cite{Alexander:2012Sci}. Solar System comet values are taken primarily from ref. \cite{Nomura:2023ASPC}, restricted to water-based measurements, with additional entries from ref. \cite{Cordiner:2025NatAs}.

\backmatter

% \bmhead{Supplementary information}

% Sample body text. Sample body text. Sample body text. Sample body text. Sample body text. Sample body text. Sample body text. Sample body text.

\bmhead{Data Availability}

The ALMA data used in this work (ADS/JAO.ALMA\#2025.A.00002.T) are publicly available through the ALMA Science Archive at \url{https://almascience.nrao.edu/aq/}.

\bmhead{Code Availability}

The radiative transfer code used in this work is publicly available at \url{https://github.com/mcordiner/sublime-d1dc}.

\bmhead{Acknowledgements}

This material is based upon work supported by the National Science Foundation under grant No. AST-2406527. T.P.C. acknowledges support from the Michigan Society of Fellows and has received funding by the Heising-Simons foundation through the 51 Pegasi  B Fellowship. Part of this research was conducted at the Jet Propulsion Laboratory, California Institute of Technology, under a contract with the National Aeronautics and Space Administration (80NM0018D0004). D.C.L. acknowledges support from the National Aeronautics and Space Administration (NASA) Astrophysics Data Analysis Program (ADAP). V.V.G. acknowledge support from the ANID -- Millennium Science Initiative Program -- Center Code NCN2024\_001, from FONDECYT Regular 1221352, and ANID CATA-BASAL project FB210003. M.A.C., N.X.R., S.N.M. and S.B.C. acknowledge support by the Planetary Science Division Internal Scientist Funding Program through the Fundamental Laboratory Research (FLaRe) work package.

This paper makes use of the following ALMA data: ADS/JAO.ALMA\#2025.A.00002.T. ALMA
is a partnership of ESO (representing its member states), NSF (USA) and NINS (Japan),
together with NRC (Canada), NSTC and ASIAA (Taiwan), and KASI (Republic of Korea), in
cooperation with the Republic of Chile. The Joint ALMA Observatory is operated by
ESO, AUI/NRAO and NAOJ.
The National Radio Astronomy Observatory is a facility of the National Science Foundation operated under cooperative agreement by Associated Universities, Inc.

\bmhead{Author contributions}

L.E.S.M., T.P.-C., and E.A.B. wrote the observing proposal on which this work is based. T.P.-C. executed the observations and performed the data reduction. L.E.S.M. performed the modeling and data analysis, and wrote the paper. L.E.S.M. and T.P.C. generated the figures. T.P.-C., M.A.C., H.W.L., D.C.L., and D.W.G. provided support with the modeling. All authors, L.E.S.M., T.P.-C., M.A.C., E.A.B., H.W.L., D.C.L., D.W.G., J.B.B., N.B., D.B.-M., D.B., S.B.C., J.C., D.F., V.V.G., S.N.M., J.W.N., A.J.R., N.X.R., and J.J.T., contributed to the evaluation and interpretation of the results and to revising the paper.

\bmhead{Competing Interests}

The authors declare no competing interest.

\onecolumn

\setcounter{figure}{0}
\setcounter{table}{0}

\renewcommand{\figurename}{Extended Data Figure}
\renewcommand{\tablename}{Extended Data Table}

\begin{figure*}
    \centering
    \includegraphics[width=1\linewidth]{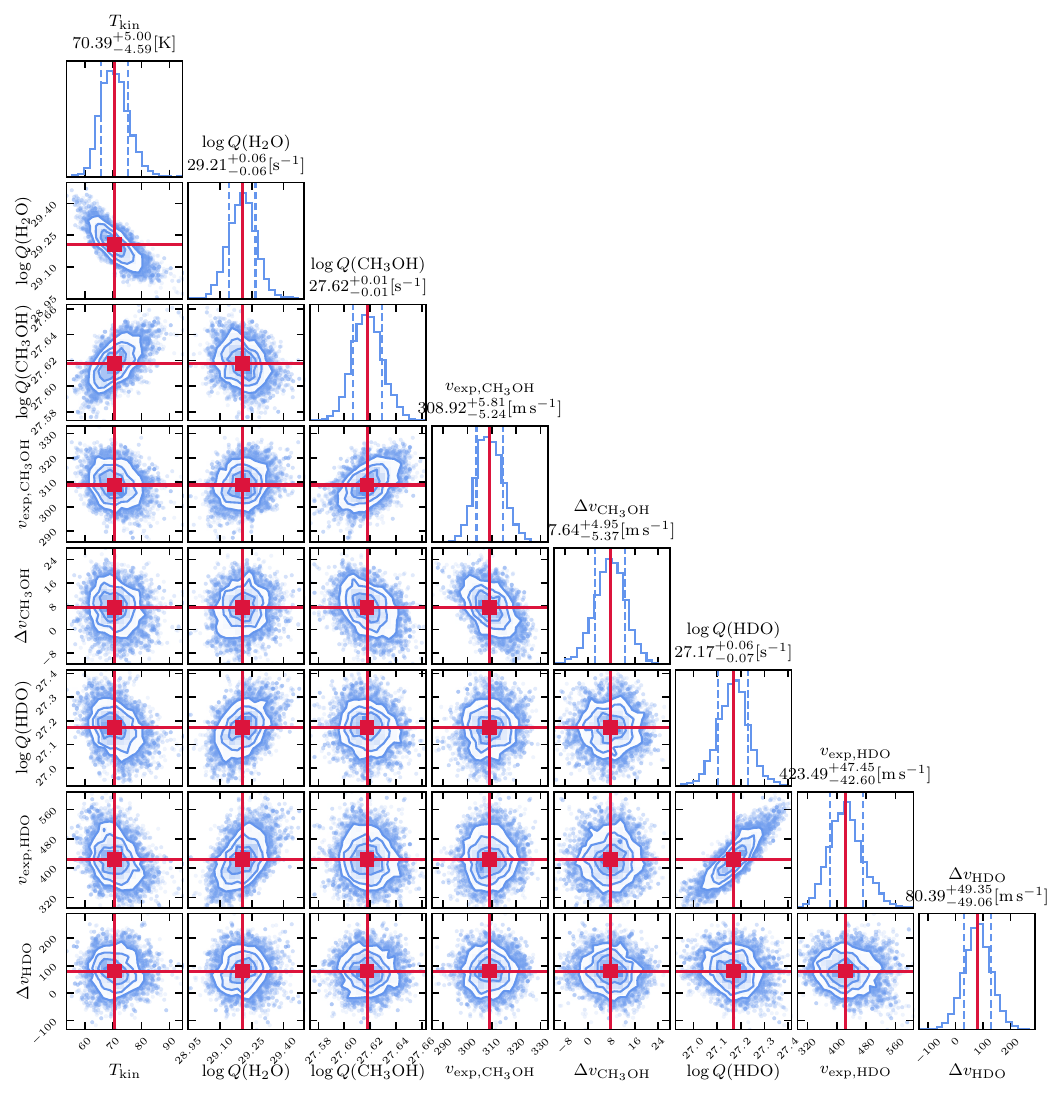}
    \caption{\textbf{Corner plot of the posterior distributions obtained from the simultaneous MCMC SUBLIME 1D fit to the HDO, H$_2$O and CH$_3$OH ALMA data}. Dashed vertical lines indicate the 16th and 84th percentiles, and the red vertical line indicates the median (50th percentile). Together, the fitted parameters provide a snapshot of the physical and chemical state of the coma.}
    \label{fig:corner_full}
\end{figure*}

\clearpage

\begin{figure*}
    \centering
    \includegraphics[width=1\linewidth]{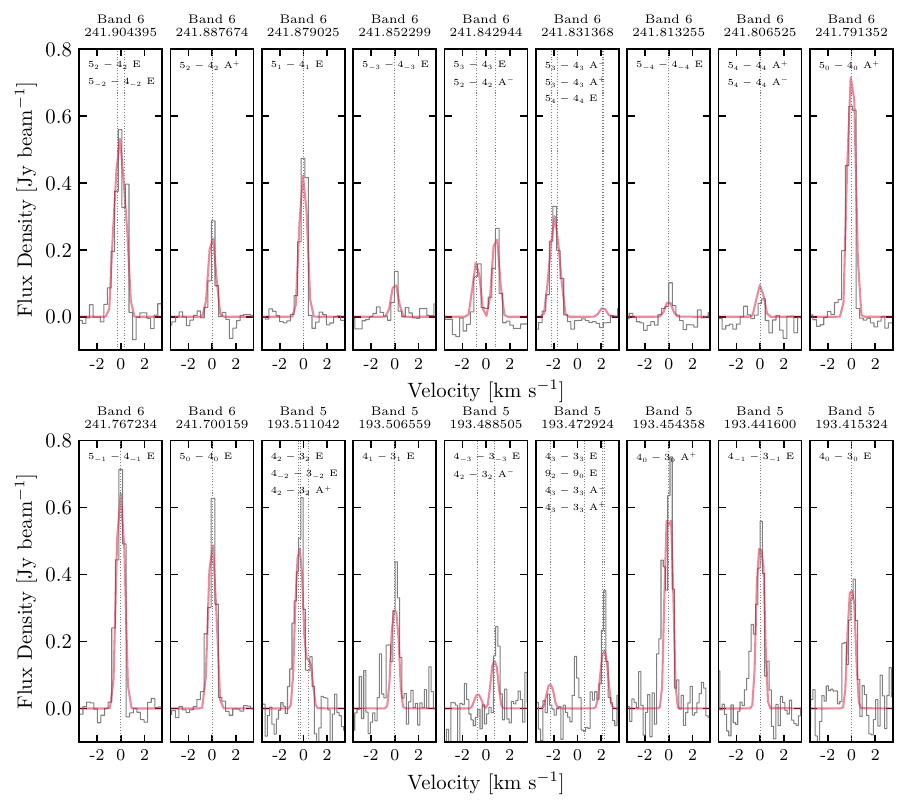}
    \caption{\textbf{CH$_3$OH Band 6 and 5 spectra compared with the best-fit MCMC SUBLIME nominal model}. All panels share the same x- and y-axis scales. For each panel, the corresponding Band and the reference frequency of the window (in GHz) are given. Vertical dotted lines mark the positions of the CH$_3$OH transitions within each window. Labels are ordered from top to bottom to match the dotted lines from left to right.}    
    \label{fig:ch3oh_spec1d}
\end{figure*}

\clearpage

\begin{figure}
    \centering
    \includegraphics[width=0.5\linewidth]{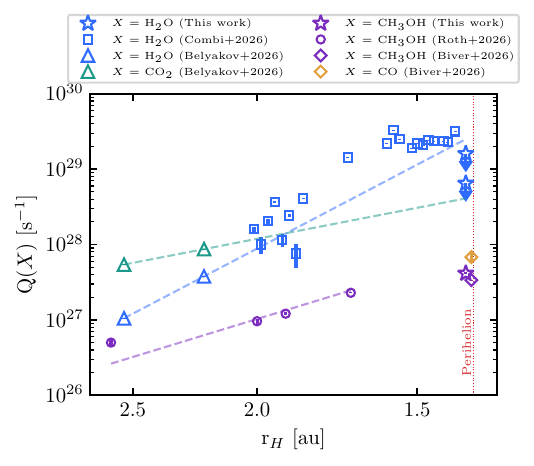}
    \caption{\textbf{Comparison of production rates derived in this work with measurements reported in the literature}. Different molecules are indicated with different colors: H$_2$O (blue), CO$_2$ (green), CH$_3$OH (purple), and CO (gold). Different datasets are indicated with different symbols: this work (stars), SOHO/SWAN \cite{Combi:2026ApJ} (squares), JWST/MIRI \cite{Belyakov:2026} (triangles), pre-perihelion ALMA \cite{Roth:2026ApJ} (circles), and IRAM 30-m \cite{Biver:2026arXiv} (diamonds). Error bars denote the quoted 1$\sigma$ uncertainties on the reported central values. The higher $Q(\mathrm{H_2O})$ upper limit is derived from the joint HDO+H$_2$O+CH$_3$OH fit, whereas the lower upper limit comes from the H$_2$O+HDO fit. The blue and green dashed lines show linear fits to the JWST/MIRI constraints for H$_2$O and CO$_2$, extrapolated to the time of our observations, which suggest $Q(\mathrm{H_2O})=Q(\mathrm{CO_2})$ at the order-of-magnitude level near $r_H\sim1.9$ au. The purple dashed line shows the fit from \cite{Roth:2026ApJ}. The only pre-perihelion measurements shown are the ALMA CH$_3$OH data from \cite{Roth:2026ApJ}.}
    \label{fig:support}
\end{figure}

\clearpage

\begin{table*}[h]
    % \centering
    % \def\arraystretch{1.3}
    % \setlength{\tabcolsep}{2pt}
    \caption{\textbf{Observational circumstances for the data used in this work (ALMA DDT 2025.A.00002.T).}}
    % \begin{tabular}{c|c|c|c|c|c|c|c|c}
    \begin{tabular}{cccccccc}
    % \hline
    % \hline
    \toprule
    \toprule    
    \multirow{2}{*}{Band} & Obs. Time &Ants.& Baselines & $\Delta$ & $r_H$ & $\theta$  & $\theta$\\
    & 4-Nov [UT] & & [m] & [au] & [au] &  [$''$] & [km] \\
    % \hline
    \midrule
    % \multirow{2}{*}{6} & 31-Oct & 13:16 - 15:50 & 8 & 10.7 - 48.9 & 2.29 & 1.354 & 5.68 $\times$ 3.98  & 9412 $\times$ 6595\\
    % & 1-Nov & 13:51 - 16:29 &8& 9.2 - 48.9 & 2.27 & 1.357 &  5.68 $\times$ 4.03 & 9363 $\times$ 6643 \\
    % & 3-Nov & 13:08 - 15:43 &8& 8.9 - 48.9 & 2.25 & 1.365 & 7.04 $\times$ 4.33  & 11492 $\times$ 7069 \\
    6 & 13:00-14:17 & 9 & 8.9 - 48.9  & 2.24 & 1.37 & 5.94$\times$4.65  & 9649$\times$7554\\
    % \hline
    5 & 14:28-15:31 & 9 & 8.9 - 48.9  & 2.24 & 1.37 & 8.81$\times$7.45 & 14312$\times$12102\\
    % & 7-Nov & 12:07 - 13:09 & 8 & 8.9 - 48.9 & 2.21 & 1.393 & 8.57 $\times$ 6.03  & 13704 $\times$ 9643 \\
    % \hline
    \botrule
    \end{tabular}
    \begin{tablenotes}
     \item $\Delta$ and $r_H$ correspond to the geocentric and heliocentric distances of 3I/ATLAS.  $\theta$ is the angular resolution (synthesized beam) of the observations in arcseconds and projected distance (km) at the geocentric distance of 3I/ATLAS. The total on-source integration time of each execution block was 50\,min for Band 6 and 20\,min for Band 5.
    \end{tablenotes}
    \label{tab:obs}
\end{table*}

\begin{table*}[h]
% \centering
\caption{\textbf{Free parameters, priors, and posterior results for the HDO+H$_2$O+CH$_3$OH MCMC retrieval.}}
\begin{tabular}{lcccc}
\toprule
\toprule
Parameter & Unit & Prior & Result\\
\midrule
% \multicolumn{3}{l}{\textbf{H$_2$O-CH$_3$OH fit}}\\
$T_{\rm kin}$ & K & $\mathcal{U}(25,150)$ & $70.4^{+5.0}_{-4.6}$ \\
$Q$(H$_2$O) & s$^{-1}$ & $\log\mathcal{U}\!\left(10^{27},10^{31}\right)$ & $\left(1.6^{+0.2}_{-0.2}\right)\times10^{29}$ \\
$Q(\mathrm{CH_3OH})$ & s$^{-1}$ & $\log\mathcal{U}\!\left(10^{21},10^{31}\right)$ & $\left(4.1^{+0.1}_{-0.1}\right)\times10^{27}$ \\
$v_{\rm exp, \, CH_3OH}$ & m s$^{-1}$ & $\mathcal{U}(50,1500)$ & $308.9^{+5.8}_{-5.2}$ \\
$\Delta v_{\rm CH_3OH}$ & m s$^{-1}$ & $\mathcal{U}(-1000,1000)$ & $7.6^{+4.9}_{-5.4}$ \\
$Q(\mathrm{HDO})$ & s$^{-1}$ & $\log\mathcal{U}\!\left(10^{21},10^{31}\right)$ & $\left(1.5^{+0.2}_{-0.2}\right)\times10^{27}$ \\
$v_{\rm exp, \, HDO}$ & m s$^{-1}$ & $\mathcal{U}(50,1500)$ & $423.5^{+47.4}_{-42.6}$ \\
$\Delta v_{\rm HDO}$ & m s$^{-1}$ & $\mathcal{U}(-1000,1000)$ & $80.4^{+49.3}_{-49.1}$ \\
\botrule
% \botrule
\end{tabular}
\begin{tablenotes}
     \item $T_\mathrm{kin}$ denotes the coma kinetic temperature. For molecule $X$, $Q(X)$ is the production rate, $v_{\mathrm{exp,}X}$ the expansion velocity, and $\Delta v_X$ the Doppler shift. $\mathcal{U}$ and $\log\mathcal{U}$ denote uniform and log-uniform priors, respectively. 
\end{tablenotes}
\label{tab:runs}
\end{table*}

% \documentclass[11pt]{article}

% \usepackage[margin=1in]{geometry}
% \usepackage{graphicx}
% \usepackage{booktabs}
% \usepackage{threeparttable}
% \usepackage{amsmath,amssymb}
% \usepackage[T1]{fontenc}

% \begin{document}

\clearpage
\vspace*{\fill}
\begin{center}
{\Large\bfseries Supplementary Information\par}
\vspace{0.6em}
{\large\bfseries Water D/H in 3I/ATLAS as a Probe of Formation Conditions in Another Planetary System\par}
\vspace{0.8em}
{\normalsize
Luis E. Salazar Manzano, Teresa Paneque-Carre\~no, Martin A. Cordiner, Edwin A. Bergin, Hsing Wen Lin, Dariusz C. Lis, David W. Gerdes, Jennifer B. Bergner, Nicolas Biver, Dominique Bockel\'ee-Morvan, Dennis Bodewits, Steven B. Charnley, Jacques Crovisier, Davide Farnocchia, Viviana V. Guzm\'an, Stefanie N. Milam, John W. Noonan, Anthony J. Remijan, Nathan X. Roth, John J. Tobin\par}
\vspace*{\fill}
\end{center}

\thispagestyle{empty}

\clearpage

\setcounter{page}{1}

% \vspace*{\fill}

\begin{center}

{\Large\bfseries Table of contents}
\end{center}

\vspace{1.5em}

\noindent
Supplementary Table 1 \hfill 2

\vspace{0.8em}

\noindent
Supplementary Figure 1 \hfill 3

\vspace{0.8em}

\noindent
Supplementary Figure 2 \hfill 4

\vspace{0.8em}

\noindent
Supplementary Figure 3 \hfill 5

\vspace{0.8em}

\noindent
Supplementary Figure 4 \hfill 6

% \vspace*{\fill}

\clearpage

\setcounter{figure}{0}
\setcounter{table}{0}

\renewcommand{\figurename}{Supplementary Figure}
\renewcommand{\tablename}{Supplementary Table}

\vspace*{\fill}

\begin{table*}[h]
\caption{\textbf{Free parameters, priors, and posterior results for the CH$_3$OH-only and H$_2$O+HDO MCMC retrievals.}}
\label{tab:restruns}
\resizebox{\textwidth}{!}{%
\begin{tabular}{lccccc} % Changed from lcccccc to lccccc since you have exactly 6 columns
\toprule
\toprule
 & & \multicolumn{2}{c}{CH$_3$OH} & \multicolumn{2}{c}{H$_2$O+HDO} \\
\cmidrule(lr){3-4} \cmidrule(lr){5-6}
Parameter & Unit & Prior & Result & Prior & Result \\
\midrule
$T_{\rm kin}$ & K & $\mathcal{U}(25,150)$ & $63.8^{+5.4}_{-4.8}$ & $\mathcal{N}(\mu=70.9,\sigma=4.9)$ & $71.3^{+4.8}_{-5.0}$ \\
$Q$(H$_2$O) & s$^{-1}$ & $\log\mathcal{U}\!\left(10^{27},10^{31}\right)$ & $\left(2.2^{+0.6}_{-0.4}\right)\times10^{29}$ & $\log\mathcal{U}\!\left(10^{27},10^{31}\right)$ & $<6.5\times10^{28}$ \\
$Q(\mathrm{CH_3OH})$ & s$^{-1}$ & $\log\mathcal{U}\!\left(10^{21},10^{31}\right)$ & $\left(4.1^{+0.1}_{-0.1}\right)\times10^{27}$ & - & - \\
$v_{\rm exp, \, CH_3OH}$ & m s$^{-1}$ & $\mathcal{U}(50,1500)$ & $312.8^{+5.9}_{-6.1}$ & - & - \\
$\Delta v_{\rm CH_3OH}$ & m s$^{-1}$ & $\mathcal{U}(-1000,1000)$ & $7.0^{+5.1}_{-5.8}$ & - & - \\
$Q(\mathrm{HDO})$ & s$^{-1}$ & - & - & $\log\mathcal{U}\!\left(10^{21},10^{31}\right)$ & $\left(8.5^{+6.0}_{-3.6}\right)\times10^{26}$ \\
$v_{\rm exp, \, HDO}$ & m s$^{-1}$ & - & - & $\mathcal{U}(50,1500)$ & $172.9^{+88.3}_{-78.3}$ \\
$\Delta v_{\rm HDO}$ & m s$^{-1}$ & - & - & $\mathcal{U}(-1000,1000)$ & $85.3^{+26.4}_{-30.3}$ \\
\bottomrule
\end{tabular}
}
\begin{tablenotes}
     \item Notation follows the same as Extended Data Table~2, with the addition that $\mathcal{N}$ means a Gaussian prior, and $\mu$ and $\sigma$ their mean and standard deviation. 
\end{tablenotes}
\end{table*}

\vspace*{\fill}

\clearpage

\begin{figure}[t]
    \centering
    \includegraphics[width=0.9\linewidth]{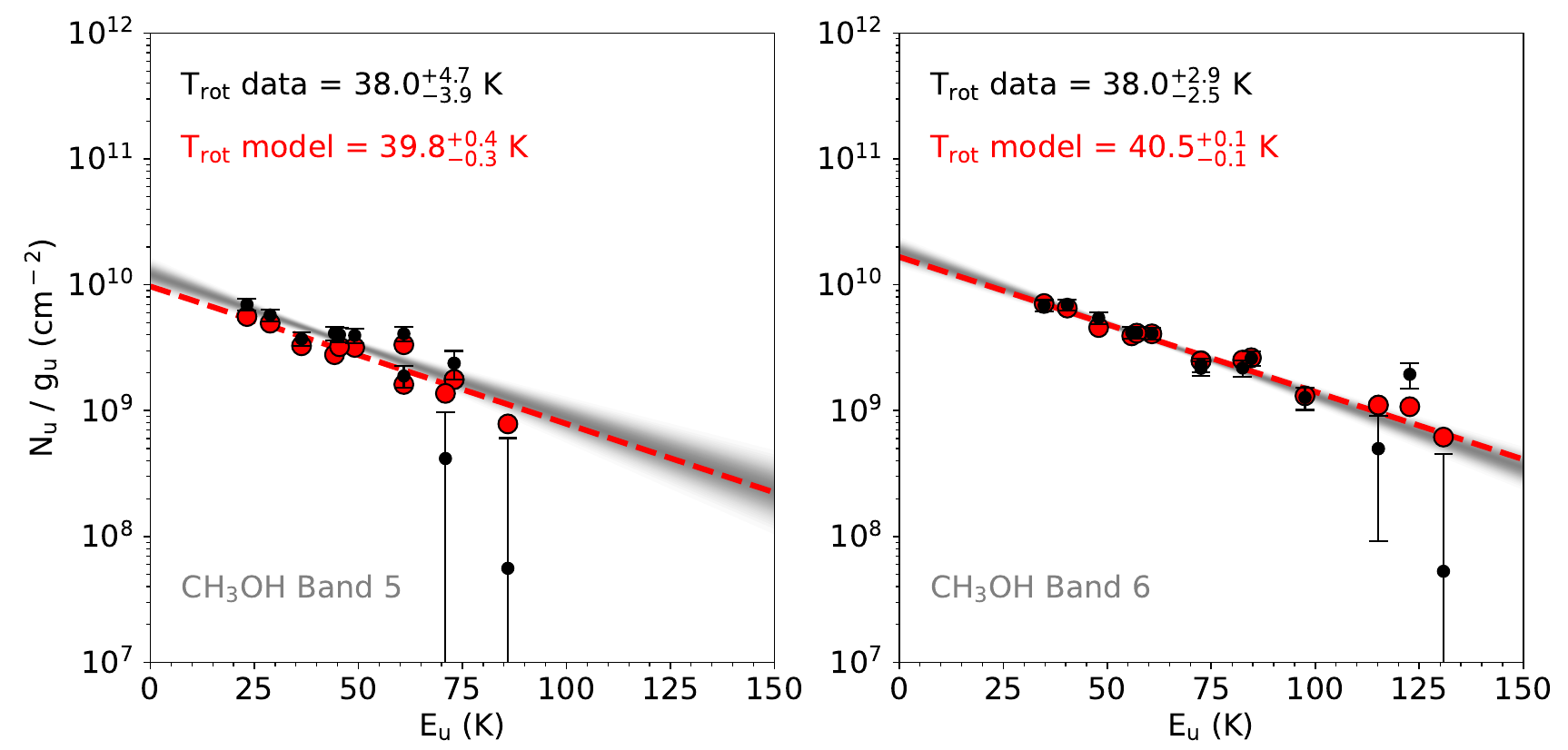}
    \caption{\textbf{Rotational diagram for the observed methanol lines, compared to the best-fit joint model (HDO+H$_2$O+CH$_3$OH)}. Left panel shows the results for Band 5 ($J = 4-3$) and right panel for Band 6 data ($J = 5-4$). Black points show the observational measurements, and the gray shaded region indicates the corresponding 1$\sigma$ uncertainties. The red circles and dashed line show the results from the model analysis.}
    \label{fig:rotation_diagram}
\end{figure}

\clearpage

\begin{figure}[t]
    \centering
    \includegraphics[width=0.49\linewidth]{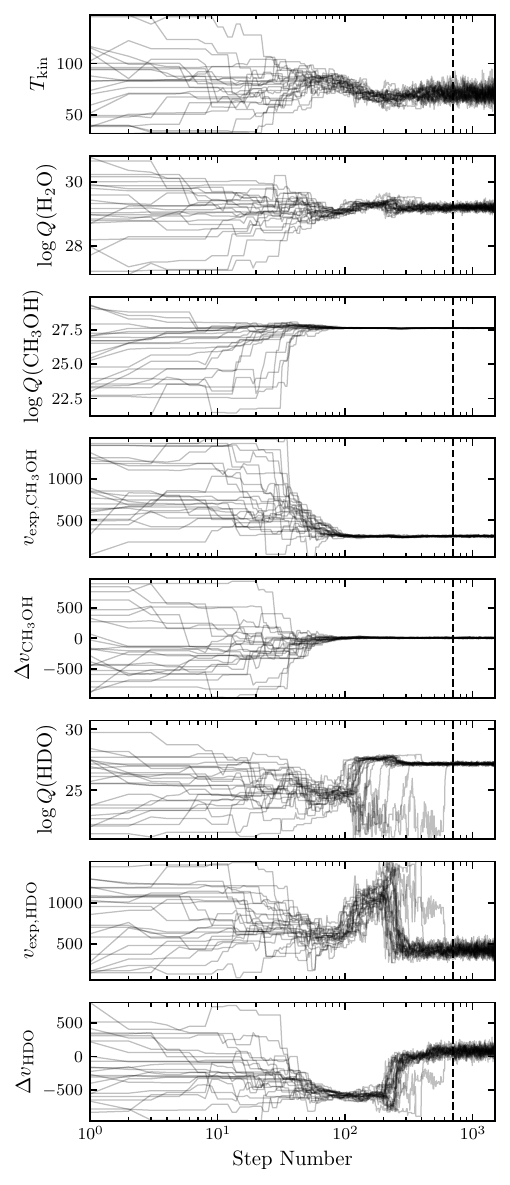}
    \caption{\textbf{Walker traces illustrating exploration and convergence of the nominal HDO+H$_2$O+CH$_3$OH MCMC retrieval}. The dashed vertical line mark the adopted burn-in cutoff. The chain includes $\sim 40000$ SUBLIME 1D model evaluations.}
    \label{fig:walkers}
\end{figure}

\clearpage

\begin{figure}[t]
    \centering
    \includegraphics[width=0.59\linewidth]{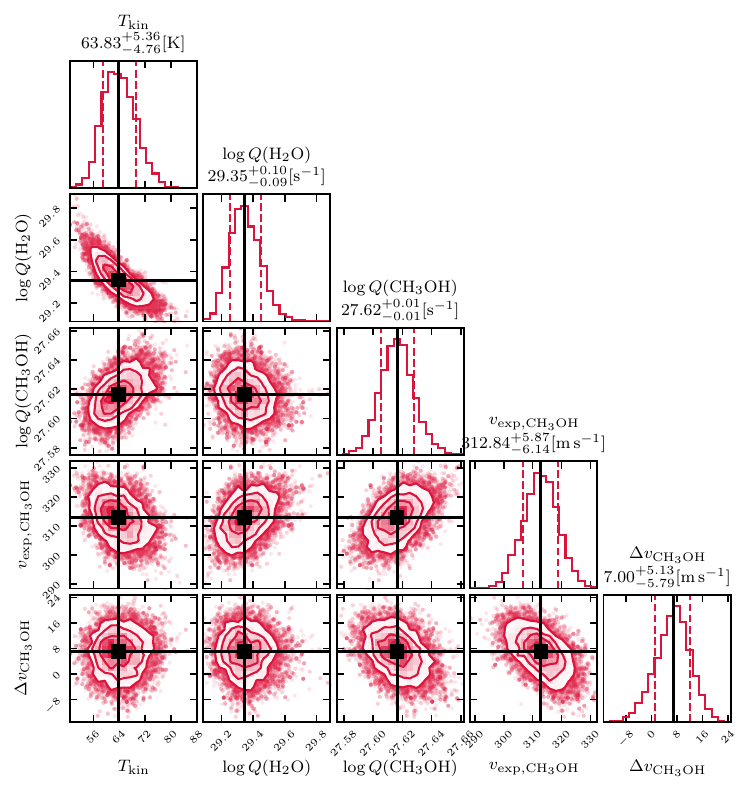}
    \includegraphics[width=0.39\linewidth]{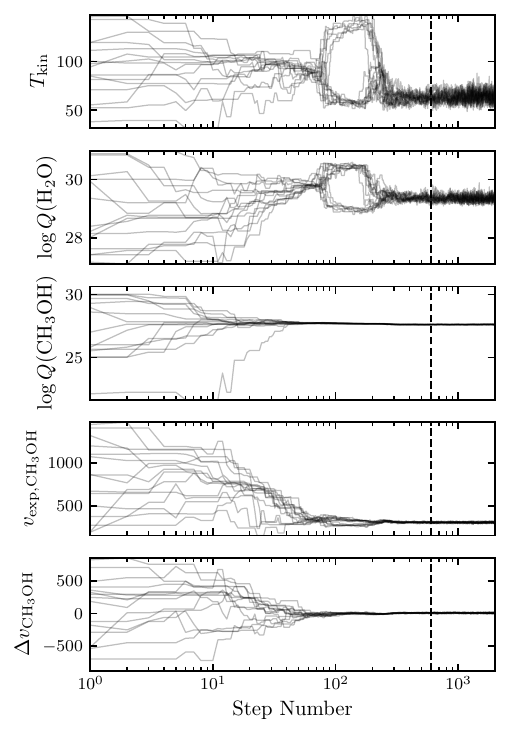}    
    \includegraphics[width=0.59\linewidth]{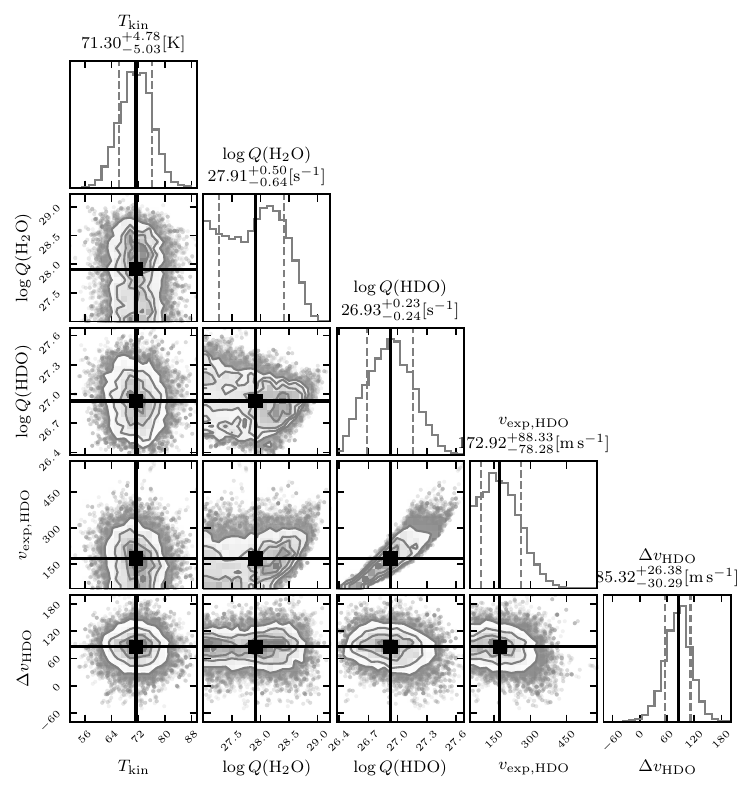}
    \includegraphics[width=0.39\linewidth]{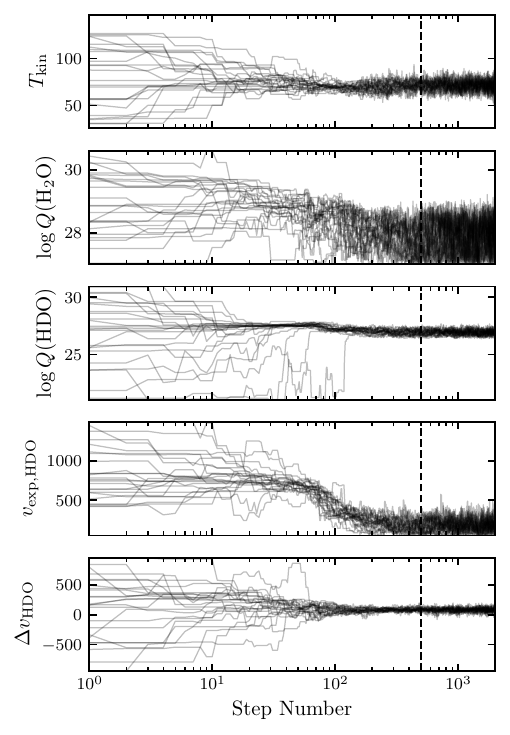}        
    \caption{\textbf{Corner plots and walker traces for additional MCMC SUBLIME 1D runs}. The top panels show the CH$_3$OH-only retrieval used to test the origin of the $Q$(H$_2$O) constraint. The bottom panels show the H$_2$O+HDO retrieval adopted as our conservative constraint on the water production rate.}
    \label{fig:restruns}
\end{figure}

\clearpage

\begin{figure}[t]
    \centering
    \includegraphics[width=1\linewidth]{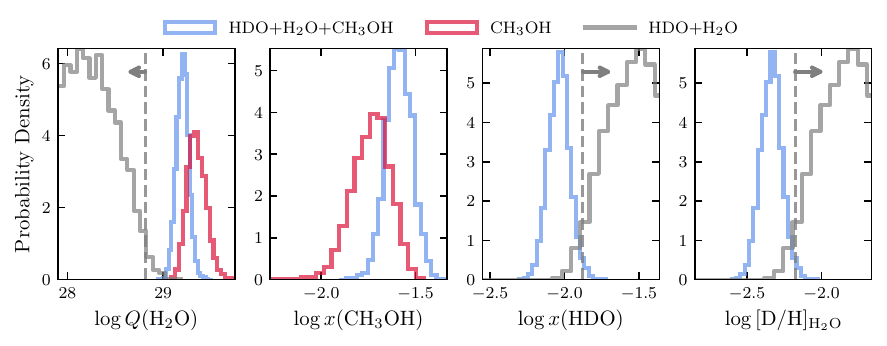}
    \caption{\textbf{Posterior distributions from our suite of MCMC SUBLIME 1D retrievals, shown for the water production rate, methanol and HDO mixing ratios, and the inferred water D/H ratio}. Blue shows the fiducial joint fit (HDO+H$_2$O+CH$_3$OH), red shows the CH$_3$OH-only fit, and gray shows the H$_2$O+HDO fit.}
    \label{fig:ratios}
\end{figure}

% \end{document}

\end{document}